\documentclass[twocolumn,aps,preprint,10pt,superscriptaddress,floatfix,showpacs]{revtex4-1}
\usepackage{graphicx}
\usepackage{epsfig,rotate,latexsym}
\usepackage{color, bm}
\usepackage{graphics}
\usepackage{dcolumn}
\usepackage{bm}
\usepackage{epsfig}
\usepackage{color}
\usepackage{epic}
\usepackage{amsmath}
\usepackage{hyperref}
\def\andname{,}

\begin{document}
\title{ Successive spin reorientations and rare-earth ordering in Nd$_{0.5}$Dy$_{0.5}$FeO$_{3}$: Experimental and \emph{ab-initio} investigations}
\author{ Ankita Singh}
\thanks{These three authors contributed equally}
\affiliation{Department of Physics, Indian Institute of Technology Roorkee, Roorkee 247 667, India
}
\author{ Sarita Rajput}
\thanks{These three authors contributed equally}
\affiliation{Department of Physics, Indian Institute of Technology Roorkee, Roorkee 247 667, India
}
\author{ Padmanabhan Balasubramanian}
\thanks{These three authors contributed equally}
\affiliation{Graphic Era University, Dehradun 248002, India}
\author{ M. Anas}
\affiliation{Department of Physics, Indian Institute of Technology Roorkee, Roorkee 247 667, India
}
\author{Francoise Damay}
\affiliation{Laboratoire L\'eon Brillouin, CEA-CNRS, CEA/Saclay, 91191 Gif sur Yvette, France
}
\author{C.M.N. Kumar}
\affiliation{Institute of Solid State Physics, Vienna University of Technology, Wiedner Hauptstr. 8-10/138, 1040 Vienna, Austria
}
\affiliation{AGH University of Science and Technology, Faculty of Physics and Applied Computer Science,  30-059 Krak\'{o}w, Poland}
\author{Gaku Eguchi}
\affiliation{Institute of Solid State Physics, Vienna University of Technology, Wiedner Hauptstr. 8-10/138, 1040 Vienna, Austria
}
\author{A. Jain}
\affiliation{Solid State Physics Division, Bhabha Atomic Research Center, Mumbai, 400 085, India
}
\affiliation {Homi Bhabha National Institute, Anushaktinagar, Mumbai 400 094, India
}
\author{S. M. Yusuf}
\affiliation{Solid State Physics Division, Bhabha Atomic Research Center, Mumbai, 400 085, India
}
\affiliation {Homi Bhabha National Institute, Anushaktinagar, Mumbai 400 094, India
}

\author{T. Maitra}
\email{tulika.maitra@ph.iitr.ac.in}
\author{V.K. Malik}
\email{vivek.malik@ph.iitr.ac.in}
\affiliation{Department of Physics, Indian Institute of Technology Roorkee, Roorkee 247 667, India
}
\date{\today}

\begin{abstract}
 In present study, the magnetic structure and spin reorientation of mixed rare-earth orthoferrite Nd$_{0.5}$Dy$_{0.5}$FeO$_3$ have been investigated. At room temperature, our neutron diffraction measurements reveal that the magnetic structure of Fe$^{3+}$ spins in Nd$_{0.5}$Dy$_{0.5}$FeO$_3$ belongs to ${\Gamma}_{4}$ irreducible representation ($G_{x}$, $F_{z}$) as observed in both parent compounds (NdFeO$_3$ and DyFeO$_3$). The neutron diffraction study also confirms the presence of spin reorientation transition where magnetic structure of Fe$^{3+}$ spins changes from ${\Gamma}_{4}$ to ${\Gamma}_{2}$($F_{x}$, $G_{z}$) representation between 75 and 20 \,K while maintaining G-type antiferromagnetic configuration. Such a gradual spin reorientation is unusual since the large single ion anisotropy of Dy$^{3+}$ ions is expected to cause an abrupt ${\Gamma}_{4}$${\rightarrow}$ ${\Gamma}_{1}$($G_{y}$) rotation of the Fe$^{3+}$ spin. Between 20 and 10 \,K, the Fe$^{3+}$ magnetic structure is  represented by ${\Gamma}_{2}$ ($F_{x}$, $G_{z}$). Unexpectedly, ${\Gamma}_{4}$ structure of Fe$^{3+}$ spins re-emerges below 10\,K  which also coincides with the development of rare-earth (Nd$^{3+}$/Dy$^{3+}$) magnetic ordering having $c_{y}^{R}$ configuration. Such re-emergence of a magnetic structure has been a rare phenomenon in orthoferrites.  The absence of a second order phase transition in rare-earth ordering, interpreted from heat capacity data, suggests the prominent role of Nd$^{3+}$-Fe$^{3+}$ and Nd$^{3+}$-Dy$^{3+}$ exchange interactions. These interactions suppress the independent rare-earth magnetic ordering observed in both parent compounds due to Nd$^{3+}$/Dy$^{3+}$-Nd$^{3+}$/Dy$^{3+}$ exchange interactions. Our density functional theory calculations including Coulomb correlation and spin-orbit interaction effects (DFT+$U$+SO) reveal that the C-type arrangement of rare-earth ions (Nd$^{3+}$/Dy$^{3+}$), with ${\Gamma}_{2}$ ($F_{x}$, $G_{z}$) configuration for Fe$^{3+}$ moments, is the ground state whereas a combination of C-type magnetic ordering (rare-earth)  and ${\Gamma}_{4}$ ($G_{x}$, $F_{z}$) configuration  of Fe$^{3+}$ spins is energetically very close to the ground state. Further, the Nd$^{3+}$-Fe$^{3+}$ and Nd$^{3+}$-Dy$^{3+}$ exchange interactions are observed to play significant roles in the complex Fe$^{3+}$ spin reorientation with the re-emergence of ${\Gamma}_{4}$ at low temperature. \\

\end{abstract}


\maketitle

\section{Introduction}
\label{intro}
Rare-earth orthoferrites $R$FeO$_{3}$ ($R$ = La, Nd, Dy,$\cdots$) have been extensively studied for their potential multiferroicity, magnetoelectric effects and other functional properties like ultrafast optical control of spins\cite{Tokunaga2008,Deng2015, Du2010,Stroppa_2010,Kimel:2004p19865,PhysRevB.84.104421,Mandal2011}. The orthoferrites belong to the family of perovskites and crystallize in the orthorhombic space group $Pbnm$, which is similar to manganites\cite{PhysRevLett.104.086402}. In comparison to highly distorted MnO$_{6}$ octahedra of the manganites, FeO$_{6}$ octahedra of the orthoferrites show least distortion with nearly equal Fe-O bond lengths at room temperature\cite{PhysRevLett.104.086402,Chiang2011,slawinski2005spin}.\par
Due to strong isotropic exchange interactions between the Fe$^{3+}$ spins, the orthoferrites have a  relatively high N\'{e}el temperature of $T_\mathrm{N1}$ ${\sim}$ 700 \,K, below which the Fe$^{3+}$ spins order in G-type antiferromagnetic configuration expressed as ${\Gamma}_{4}$ ($G_{x}$,  $F_{z}$) irreducible representation\cite{yamaguchi1974}. Here, $G_{x}$ represents G-type antiferromagnetic ordering of Fe$^{3+}$ spins with moment direction aligned along crystallographic $a$ direction which is the easy anisotropy axis and $F_{z}$ corresponds to the weak ferromagnetic component arising from the canting of the Fe$^{3+}$ spins due to Dzyloshinski-Moriya interactions. Isotropic exchange interaction between $R^{3+}$ and Fe$^{3+}$ ions polarizes $R^{3+}$ moments. As temperature decreases, the polarization of $R^{3+}$ moments increases.   Due to antisymmetric, and the anisotropic-symmetric  parts of exchange interactions between $R^{3+}$ and Fe$^{3+}$ ions, the easy axis of Fe$^{3+}$ spins undergo spin-reorientation from the $a$ axis to either $b$ or $c$ axis of the crystal depending upon the type of rare-earth ion\cite{yamaguchi1974}. In general, the orthoferrites show a gradual rotation to $c$ axis with temperature as seen in $R'$FeO$_{3}$($R'$=Nd, Tb, Er)\cite{Yuan2013, Gorodetsky1973} or an abrupt one near 50 \,K as observed in DyFeO$_{3}$\cite{Wang2016, yamaguchi1974}.\par 
    Among the orthoferrites, the most studied compound NdFeO$_{3}$ has a N\'{e}el temperature ($T_\mathrm{N1}$) of 690\,K\cite{slawinski2005spin}. Below $T_\mathrm{N1}$, Fe$^{3+}$ magnetic moments order in antiferromagnetic configuration and the magnetic structure of Fe$^{3+}$ sublattice is represented by ${\Gamma}_{4}$ ($G_{x}$, $F_{z}$) \cite{Yuan2013}. As the temperature is reduced below 200\,K, the Fe$^{3+}$ spins continuously rotate in the $ac$ plane between 200\,K and 105\,K thereby resulting in ${\Gamma}_{2}$($G_z$,$F_x$) magnetic structure below 105\,K. 
Recent terahertz spectroscopy studies in single crystals of NdFeO$_{3}$ reveal a correlation between the gap in spin-wave dispersion and the changes in magnetic anisotropy of the system in the spin reorientation region\cite{Constable2014}.
Below 25\,K, additional magnetic Bragg peak (100), corresponding to the ${\Gamma}_{2}$ ($c^R_y,f^R_y)$ representation, develops in neutron diffraction, indicating polarization of Nd$^{3+}$ magnetic moments\cite{bartolome1997single}.  Nd$^{3+}$-Fe$^{3+}$ exchange interaction is responsible inducing such a polarization in Nd$^{3+}$ magnetic moments. The weak ferromagnetic moments ($f^R_y$) associated with Nd$^{3+}$ polarization are antiparallel to the weak ferromagnetic moments of Fe$^{3+}$($F_x$) and compensate each other at 8\,K. The ordering of Nd$^{3+}$ and Fe$^{3+}$ moments are compatible with each other from the group theory. 
%
Below $T_\mathrm{N2}$=1.05\,K, the Nd$^{3+}$-Nd$^{3+}$ interactions give rise to a proper long range independent G-type antiferromagnetic ordering of Nd$^{3+}$ moments\cite{bartolome1997single}. The observation of such a long range antiferromagnetic ordering is confirmed by observation of a sharp peak in specific heat data which is superimposed on the low temperature tail of Schottky anomaly\cite{Bartolome1994,Pataud1970}.
\par
     The most intriguing orthoferrite happens to be DyFeO$_{3}$. Though isostructural with NdFeO$_3$, the degree of structural distortion in DyFeO$_{3}$ is greater. Like all orthoferrites, the Fe$^{3+}$ spins of DyFeO$_3$ also order in antiferromagnetic  configuration expressed by ${\Gamma}_{4}$ ($G_{x}$, $F_{z}$) representation below $T_\mathrm{N1}$(${\sim}$650\,K)\cite{Gorodetsky1968}. However, near 50 \,K,  an abrupt spin reorientation occurs, as a result, the magnetic structure changes from ${\Gamma}_{4}$ ($G_{x}$, $F_{z}$) to ${\Gamma}_{1}$($G_{y}$) representation which does not have any ferromagnetic component. This is also known as Morin transition. The ${\Gamma}_{1}$ ($G_{y}$) structure persists till the lowest temperatures at zero field\cite{Prelorendjo1980}.
 Below $T_\mathrm{N2}$ ${\sim}$ 4\,K, a long range independent antiferromagnetic ordering of Dy$^{3+}$ moments is observed with ${\Gamma}_{5}$($g_{x}^{R}$,$a_{y}^{R}$) configuration making an angle of 33$^{\circ}$ with the $b$ axis while the moments are confined to the $ab$ plane. The studies on similar compounds viz. DyAlO$_{3}$ and DyScO$_{3}$, have also shown that the Dy$^{3+}$ moments order in the same ${\Gamma}_{5}$($g_{x}^{R}$,$a_{y}^{R}$) representation. This type of magnetic ordering is mainly dictated by the  Dy$^{3+}$-Dy$^{3+}$ exchange interactions. 
Studies have thus revealed that the strong Dy$^{3+}$ single ion anisotropy determines the ground state magnetic properties\cite{Holmes1972, Wu2017}. Recent reports suggest that below $T_\mathrm{N2}$, the Fe$^{3+}$ spins orient into a different magnetic structure due to Dy$^{3+}$-Fe$^{3+}$ interaction\cite{Zhao2014}.
In DyFeO$_{3}$, the ${\Gamma}_{5}$ structure is not symmetry compatible with ${\Gamma}_{1}$. The overall reduction in symmetry due to the Dy$^{3+}$ ordering results in linear magneto-electric effect\cite{Yamaguchi1973}. A large ferroelectric polarization is also observed due to the weak ferromagnetic ordering induced by magnetic field along the $c$ axis of the DyFeO$_{3}$ crystals\cite{Tokunaga2008}.
\par

          Substitution at the $R$ or Fe site provide the possibility of achieving the tunability of desired electrical and magnetic properties in orthoferrites. Additionally, exchange interactions and magnetic anisotropies can be tuned to observe successive spin reorientations which can be exploited for magnetic switching at different temperature regimes. For instance, substitution of Fe by Mn in NdFeO$_{3}$ affects the spin reorientation such that the ${\Gamma}_{1}$ magnetic structure occurs in a wide temperature range suppressing the ${\Gamma}_{4}$ magnetic structure.\cite{Ankita2017}. In YFe$_{1-x}$Mn$_{x}$O$_{3}$, ferroelectricity and magnetodielectric effects are observed\cite{Mandal2011}. 
Similarly, presence of two rare-earth ions at the $R$ site also yield an interesting set of properties. For instance, in Dy$_{0.7}$Pr$_{0.3}$FeO$_{3}$ an enhanced ferroelectricity is observed\cite{Tokunaga2014}. In Dy$_{0.5}$Ho$_{0.5}$FeO$_{3}$, though the Morin transition occurs at 45\,K, an additional ${\Gamma}_{1}$${\rightarrow}$${\Gamma}_{2}$ transition occurs at 25\,K\cite{Belov1999}.
Interestingly, in Dy$_{0.5}$Pr$_{0.5}$FeO$_{3}$, the Morin transition occurs at 75\,K itself, followed by additional reorientation at lower temperature\cite{Wu2014}. Thus the Morin transition seems unaffected by Ho$^{3+}$, while greatly enhanced by Pr$^{3+}$ ion, even though the low temperature rare-earth ordering is affected. Surprisingly,  ${\Gamma}_{4}$${\rightarrow}$${\Gamma}_{2}$ reorientation occurs at much lower temperature (10-6\,K) in PrFeO$_3$\cite{Li2019}.
Thus, the nature of exchange interactions between the two different rare-earth ions and also with that of Fe$^{3+}$ can lead to behaviour different from that of both parent compounds.  
%
\par
  In view of this, it would be interesting to observe behavior of the compound Nd$_{0.5}$Dy$_{0.5}$FeO$_{3}$(NDFO). Nd$^{3+}$, being a Kramer's ion similar to Dy$^{3+}$, the result can be more complex. In the present work, the magnetic and thermodynamic properties of NDFO have been investigated using the  bulk magnetization, neutron diffraction, and specific heat techniques.  In addition, density functional theory calculations have been used to establish the ground state magnetic structure of NDFO as well as to qualitatively understand the underlying mechanism of spin reorientation and the nature of rare-earth ordering. Our results show that the Nd$^{3+}$-Fe$^{3+}$ interactions result in ${\Gamma}_{4}$${\rightarrow}$${\Gamma}_{2}$ reorientation at relatively higher temperature (75\,K) in comparison to Morin transition.
%
The stronger Dy$^{3+}$-Dy$^{3+}$ interactions which result in rare-earth ordering and the strong single ion anisotropy of Dy$^{3+}$ can compete with the Nd$^{3+}$-Fe$^{3+}$ exchange interactions, thereby affecting the Morin transition and also the rare-earth ordering. Surprisingly, the presence of 50${\%}$ Nd completely suppresses the Morin transition, yielding a successive two fold spin reorientation (${\Gamma}_{4}$${\rightarrow}$${\Gamma}_{2}$${\rightarrow}$${\Gamma}_{4}$) of the Fe$^{3+}$ spins. Also, the independent rare-earth ordering completely vanishes even down to 0.4\,K as seen in our specific heat studies. Rather, the Nd$^{3+}$/Dy$^{3+}$ moments order by means of induced polarization by $R^{3+}$-Fe$^{3+}$ interactions. Also, the absence of any magnetoelectric effect as predicted by symmetry is confirmed by dielectric studies under magnetic field. 

%
%
%
%
\section{Methods}
\label{methd}
\subsection{Experimental}
\label{Expmethd}
Powder sample of NDFO was synthesized using solid state reaction method. Nd$_{2}$O$_{3}$, Dy$_{2}$O$_{3}$, and Fe$_2$O$_3$ were weighed according to appropriate stoichiometry and grounded in an agate mortar for 12 hours. The sample was sintered at consecutively 1200$^\circ$C, 1300$^\circ$C, and 1400$^\circ$C for 24 hours with intermediate grindings. Structural phase of the sample was  identified using a Bruker D8 two circle x-ray diffractometer at Cu $K_{\alpha}$ wavelength. 
Bulk magnetization measurements were performed using Superconducting Quantum User interface device (SQUID) magnetometer of Quantum Design Inc.'s  magnetic measurement system-XL (MPMS-XL) and vibrating sample magnetometer option of Quantum Design Inc.'s Dynacool physical properties measurement system (PPMS). 
 Temperature dependent zero field cooled (ZFC) and field cooled (FC) measurements were carried out to identify the different magnetic transitions and their respective transition temperatures. Field variation of magnetization was carried out at various temperatures between 300 and 5\,K. 
%
%
Heat capacity measurements were carried out using heat capacity option of Quantum design Inc.'s PPMS for temperature range of 2-200K, and $^3$He refrigerator for temperature range of 0.4-20\,K. Magneto-dielectric measurements were carried out using a custom design probe for PPMS sample sample chamber.\par
Neutron diffraction studies in zero magnetic field were carried out at various temperatures in the range of 1.5-300\,K to identify the crystal as well as magnetic structure and their variations as a function of temperature. The  neutron diffraction measurements were performed at powder diffractometers G 4-1 
(${\lambda}$ = 2.4206\,${\mathrm{\AA}}$) of Laboratoire L\'eon Brillouin (LLB), Saclay, France.  The diffraction data were analyzed using FullProf \cite{rodriguez1990fullprof} suite of programs utilizing the Rietveld refinement method \cite{rietveld1969profile}. Magnetic structure was determined using the irreducible representations from BasIreps \cite{Hovestreydt:wi0099} and refined using FullProf. 
\subsection{Theoretical}
\label{Theomethd}
Electronic structure of NDFO was obtained using the projector-augmented wave (PAW) psuedopotential and a plane wave basis method within the density functional theory framework as implemented in the Vienna Ab-initio simulation program (VASP)\cite{kresse1996efficient}. Calculations were performed within Perdew-Burke-Ernzerhof generalized gradient approximation (PBE-GGA)\cite{perdew1996generalized} and GGA+$U$ approximation \cite{anisimov}. The structure was relaxed keeping the Nd/Dy $4f$ states as frozen in the core. Ionic positions were relaxed until the forces on the ions are less than 0.1\,meV$\mathrm{\AA}^{-1}$. For the subsequent self-consistent calculations, the Nd/Dy $4f$ states were treated as valence states.
 The (Fe) $3d, 4s$, O $2s, 2p$ and Nd $5p, 5d, 6s$ states were treated as valence states. An energy cut-off of 450\,eV was used for the plane wave basis set while a 6$\times$6$\times$6 Monkhorst-Pack $k$-mesh centered at ${\Gamma}$ point was used for performing the Brillouin zone integrations.  
\begin{figure}[h!] \center
       \begin{picture}(190,155)
      \put(-35,-60){\includegraphics[height=220\unitlength,width=250\unitlength]{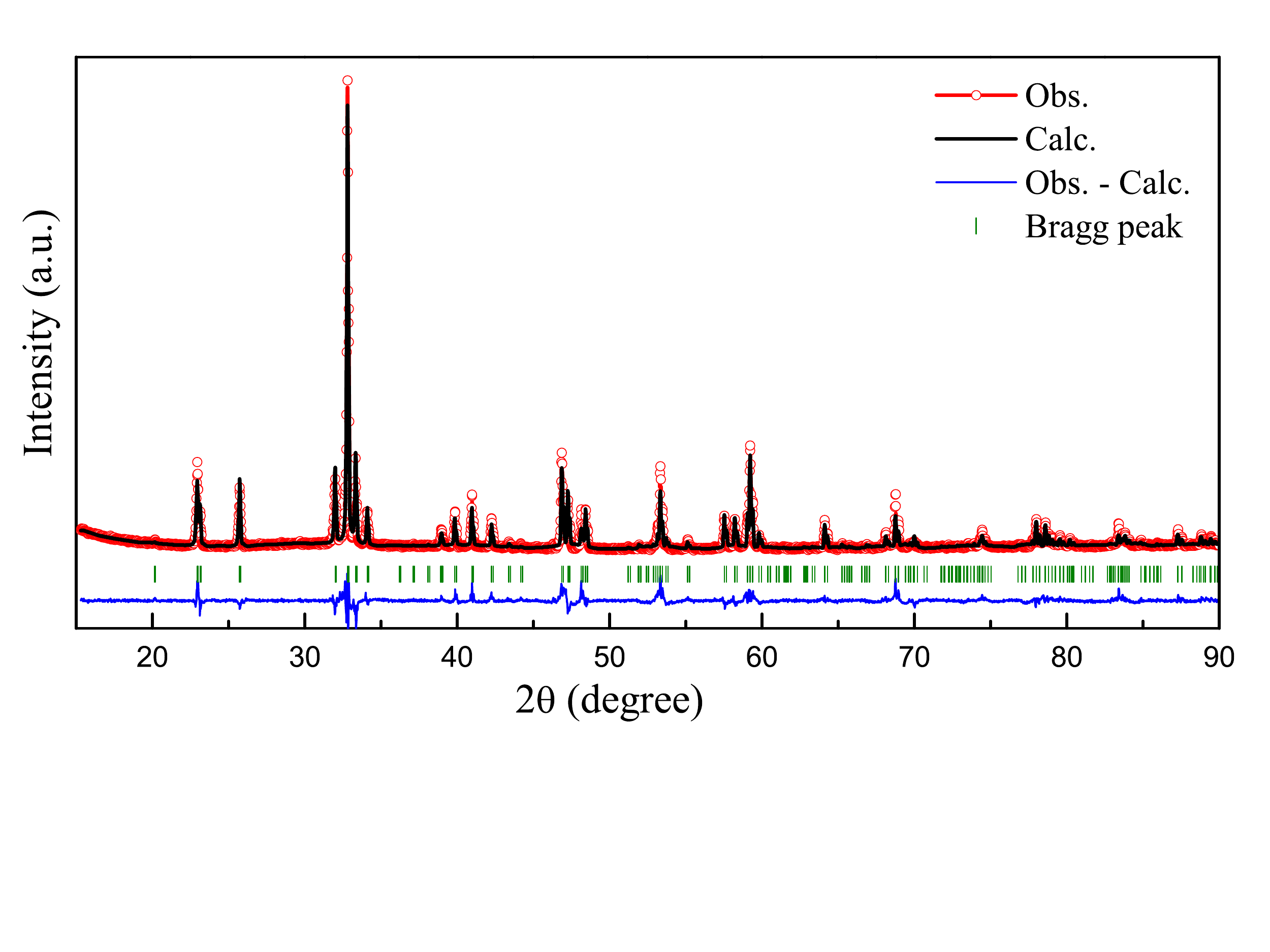}}
      \end{picture}
\caption{(color online) Observed and simulated x-ray diffraction pattern of NDFO at 300 K, refined using $Pbnm$ space group.}
\label{xrd}
\end{figure}

\begin{figure}[h!] \center
       \begin{picture}(240,170)
        \put(-10,-5){\includegraphics[width=260\unitlength,]{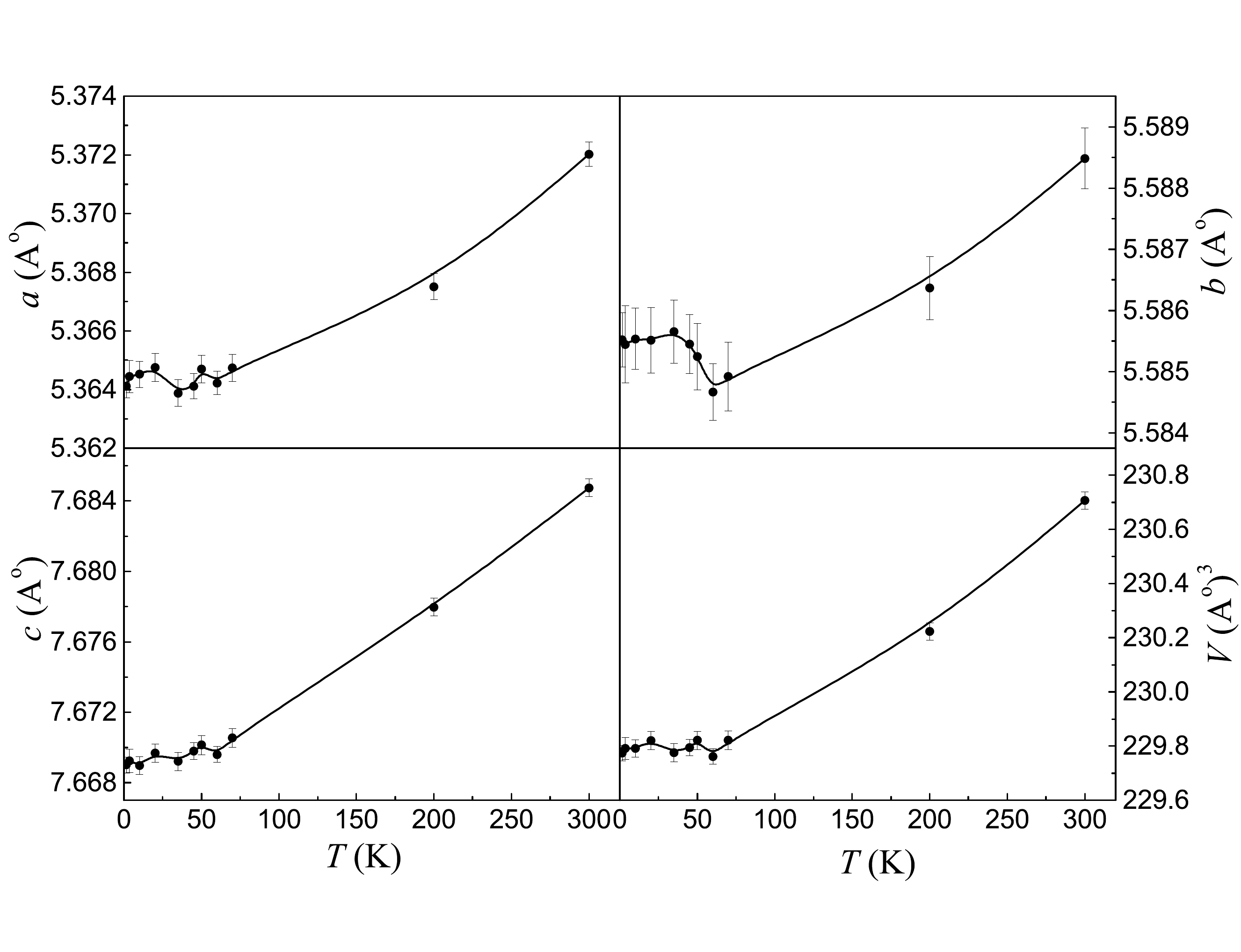}}
       \end{picture}
\caption {Temperature variation of lattice parameters and unit cell volume of NDFO. Solid line is guide to eye}
\label{latticeparams}
\end{figure}
\section{Experimental results}
\label{res}
\subsection{Structural Characterization}
\label{stuct_char}
Fig.~\ref{xrd} shows the room temperature powder x-ray diffraction pattern of the NDFO. The compound crystallizes in the orthorhombic $Pbnm$ space group. The observed neutron diffraction pattern of NDFO shows no traces of any impurity phase.\par
The neutron diffraction patterns were collected systematically at regular intervals between 1.5\,K to  300\,K. In Table~\ref{Table_1}, the lattice constants along with Fe-O($M$ = Fe) and Nd(Dy)-O bond lengths of NDFO, obtained form the combined refinement of the x-ray and neutron diffraction data at 300 K are shown. The values of bond lengths and bond angles at 1.5 K have been obtained from the refinement of the  neutron diffraction data. Three different Fe-O bond lengths correspond to the apical and in-plane bond lengths. Unlike the manganites which have highly unequal bond lengths, the orthoferrites have nearly equal Fe-O bond lengths. From Table~\ref{Table_1}, we can infer that a contraction in out-of plane bond length occurs, while the in-plane bond lengths remain nearly constant with temperature.\par
%
\begin{table}
\caption{Structural parameters of Nd$_{0.5}$Dy$_{0.5}$FeO$_{3}$ obtained using combined Retvield refinement of neutron \& x-ray diffraction data at 300\,K and refinement of neutron diffraction data at 1.5\,K.}
\begin{tabular}{p{2.5cm}p{2.0cm} p{2.0cm} c c c c c }\hline\hline 
 & \multicolumn{3}{c}{} & \multicolumn{1}{c}{} & \multicolumn{1}{c}{}& \multicolumn{1}{c}{}\\
Parameters & 300\,K &&  1.5\,K   \\& (Neutron &&(Neutron)&&&&\\&  \& X-ray)&&&&&&\\  \hline
$a$({\AA}) & 5.377(2) && 5.364(2)  \\
$b$({\AA}) &  5.597(2) && 5.586(3) \\  
$c$({\AA}) & 7.694(4) && 7.669(2) \\ 
Fe-O(1)(m){\AA} & 2.021(3)& & 1.966(4) \\  
Fe-O(2)(l){\AA} & 2.040(2) &&  2.018(2)  \\
Fe-O(2)(s){\AA} & 2.027(2) && 2.037(2) \\ 
\hline\hline
\end{tabular}
\label{Table_1}
\end{table}
The lattice parameters of NDFO are intermediate to the lattice parameters of both the parent compounds (NdFeO$_{3}$ and DyFeO$_{3}$)\cite{slawinski2005spin, Wang2016}. 
The ratio $b$/$a$, which increases monotonically with the atomic number of the rare-earth, has a value of 1.024 in case of NdFeO$_{3}$. In NDFO, the value of $b$/$a$ is 1.040  which is close to the value obtained for DyFeO$_{3}$\cite{slawinski2005spin,Wang2016}.
 In Fig.~\ref{latticeparams}, the temperature variation of lattice parameters and unit cell volume of NDFO is shown. Starting from 300\,K, ${\it a}$, ${\it b}$ and ${\it c}$ show a sharper and continuous decrease till 60\,K. 
From 60\, K till 30\, K, $b$ shows an increase indicating a crossover from positive to negative thermal expansion. Below 30\, K, $b$ shows a plateau like behaviour with no significant change with further decrease in temperature. Similar variation in $b$ is also reported for NdFeO$_{3}$\cite{slawinski2005spin}. However in NdFeO$_{3}$, the rise in $b$ starts at 150\,K itself which is more gradual and finally converging to a plateau near 50\,K. This temperature range (150-50\,K) also coincide with the spin reorientation range of NdFeO$_3$. The trend reversal of $b$ below 60\, K is also accompanied by slope changes in $a$ and $c$. The variation of $c$ below 60 \,K is considerably smaller compared to the fluctuations in $a$. The slope change in $b$ coincides with the advent of spin reorientation of the Fe$^{3+}$ spins in NDFO, which is discussed in detail further in the subsequent sections. The variation of unit cell volume $V$ (Fig.~\ref{latticeparams}(d)) is similar to reported variation in NdFeO$_{3}$\cite{slawinski2005spin}. The unit cell volume variation does not show any signs related to magneto-elastic or magneto-volume effect due to spin reorientation.\par
\begin{figure}[b!] \center
       \begin{picture}(330,160)
        \put(-5,-15){\includegraphics[width=270\unitlength,]{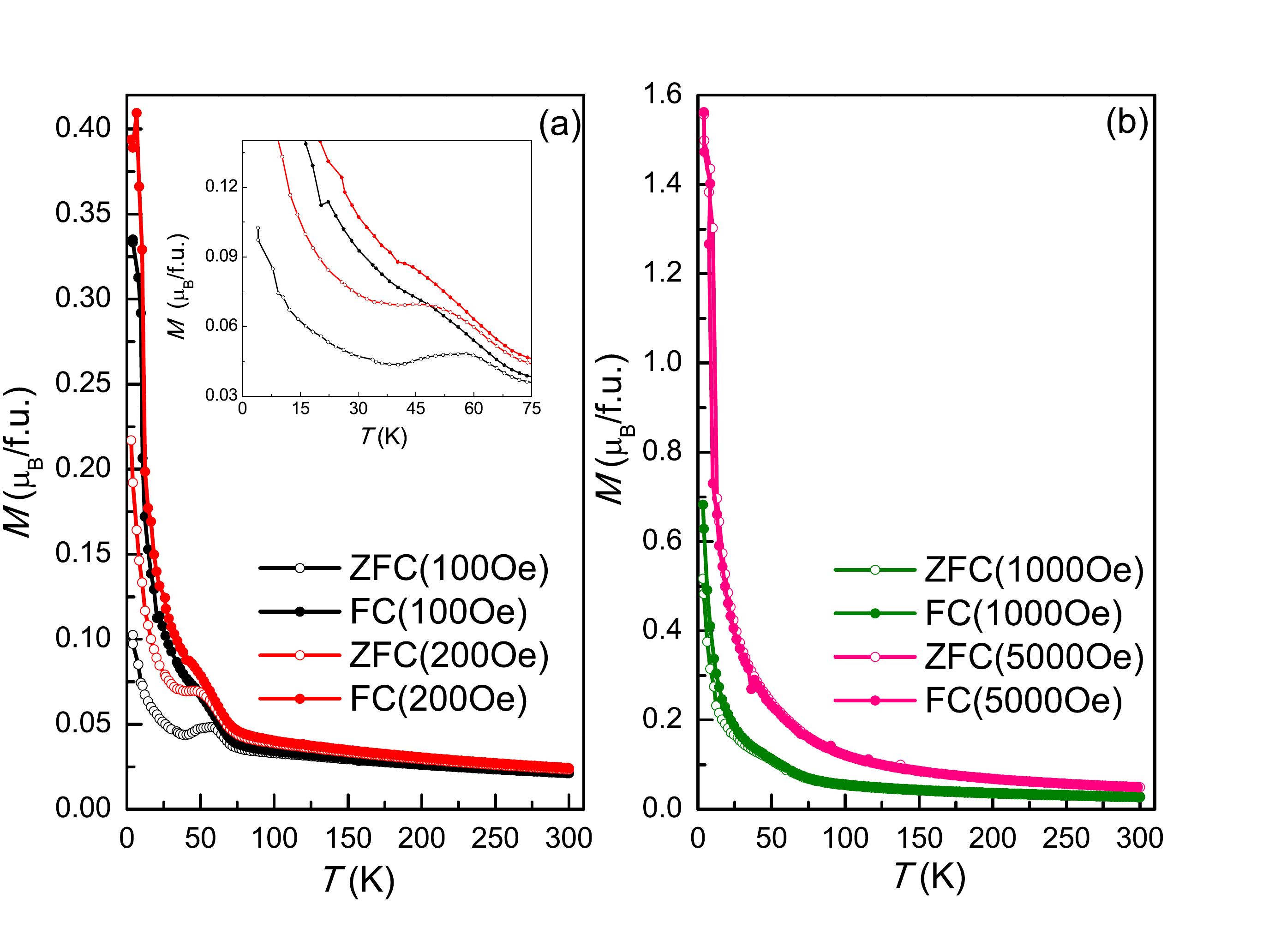}}
       \end{picture}
\caption{(color online)  ZFC-FC plots of NDFO at (a) 100 Oe and (b) 1000 and 5000 Oe.}
\label{ZFC-FC}
\end{figure}

\subsection{Magnetic properties}
\subsubsection{DC magnetization}
\label{avg_mag}
As shown in Fig.~\ref{ZFC-FC}, zero field cooled (ZFC) and field cooled (FC) magnetization measurements were performed from 2\,K to 300\,K in presence of 100, 200, 500, 1000, and 5000\,Oe (external magnetic field values). 
Both parent compounds, NdFeO$_3$ and DyFeO$_3$, undergo transition from paramagnetic state to G-type antiferromagnetic state with magnetic structure represented by ${\Gamma}_{4}$($G_{x}$, $F_{z}$) at 690\,K and 650\,K, respectively\cite{slawinski2005spin,Gorodetsky1968}. 
Thus at 300\,K, NDFO is already in the  antiferromagnetically ordered state with ($G_{x}$, $F_{z}$) structure, similar to end compounds. 
 As shown in Fig.~\ref{ZFC-FC}a, a magnetic transition is observed in the ZFC-FC magnetization data below $\approx$ 75\,K. In DyFeO$_{3}$, an abrupt spin reorientation is observed at 50\,K which corresponds to ${\Gamma}_{4}{\rightarrow}{\Gamma}_{1}$ transition i.e Morin Transition. The ${\Gamma}_{1}$ representation, which corresponds to  antiferromagnetic $G_{y}$  magnetic structure in Bertaut notation, does not contain any net effective  moment. Although low field (100\,Oe and 200\,Oe) ZFC measurements show a drop in magnetization below 60\,K, an increase in 100\,Oe and 200\,Oe  FC magnetization is observed contrary to the expected behavior from ${\Gamma}_{1}$ ($G_{y}$) magnetic structure.  The ZFC magnetization values also increase  below 35\,K during 100 and 200\,Oe ZFC measurements. As shown in Fig.~\ref{ZFC-FC}(a), an observed  bifurcation of ZFC and FC magnetization values can also be expected from a spin glass system, but ac susceptibility measurements (not shown) on NDFO rule out the possibility of a spin glass phase. 
As shown in Fig.~\ref{ZFC-FC}(b),  ZFC-FC measurements were also carried out at higher magnetic fields of 1000 and 5000\, Oe. With increasing field, we see a gradual suppression of the transition below 75\,K. As temperature decreases,  at a field of 5000\,Oe, the ZFC as well as FC magnetization show a continuous increase without any noticeable bifurcation.\par
 In both parent compounds (NdFeO$_3$ and DyFeO$_3$), the application of external magnetic fields have drastically different effects.
In NdFeO$_{3}$, although the magnetization data under applied magnetic fields show anisotropic behavior and magnetization reversal, but it is not known to cause a field-induced spin reorientation\cite{Yuan2013}.
However in DyFeO$_{3}$, though the Dy$^{3+}$-Fe$^{3+}$ exchange interactions do not effect independent ordering of both the ions, application of an external magnetic field along various crystal axes result in multiple spin reorientations. The reorientation can be spin-flop as well as screw rotations depending upon the direction of magnetic field. 
Thus near 77\,K (well above the Morin transition), at a critical field of 4.5\,T along the $a$ direction the Fe$^{3+}$ spins in DyFeO$_{3}$ undergo a field-induced ${\Gamma}_{4}$${\rightarrow}$${\Gamma}_{2}$ reorientation\cite{Prelorendjo1980}. 
For field along $b$ and $c$ directions, the ${\Gamma}_{4}$ magnetic structure is retained and eventually, the Morin transition is suppressed. 
This suggests that, in NDFO the "$x$" component of the molecular field ${\bf H_{Nd-Fe}}$ can also cause the reorientation of a significant fraction of the total Fe$^{3+}$ spins, in a manner similar to that of the applied field in DyFeO$_{3}$.
Thus in NDFO, a continuous ${\Gamma}_{4}$${\rightarrow}$${\Gamma}_{2}$ reorientation occurs below 75\,K, eventually suppressing the Morin transition. 
%
\par
  \begin{figure}[t!] \center
       \begin{picture}(250,180)
        \put(-35,-10){\includegraphics[height=210\unitlength,width=315\unitlength]{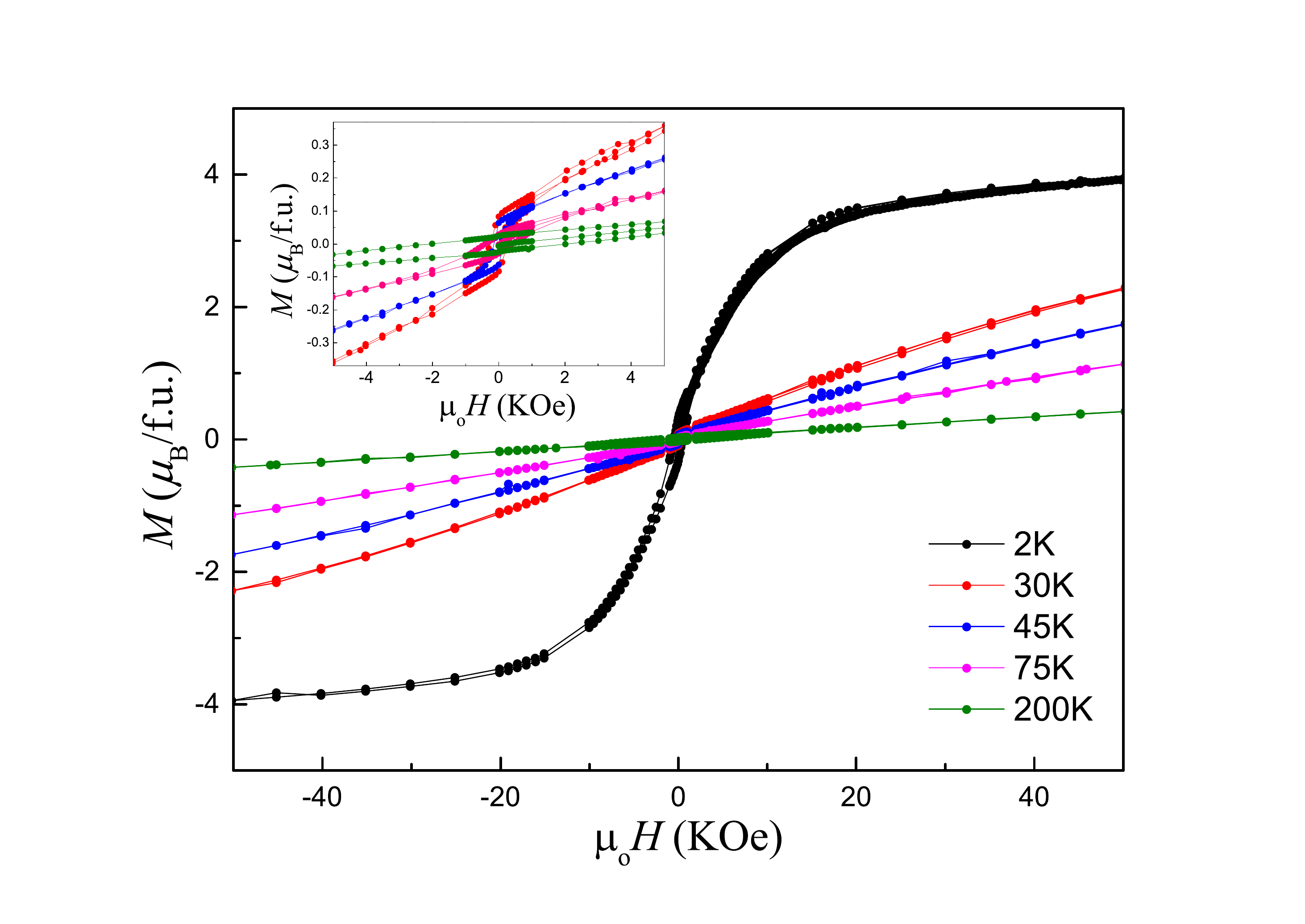}}
    
       \end{picture}
        \caption{(color online) M-H plots of NDFO at 200\,K, 75\,K,  45\,K, 30\,K,  and  2\,K.}
        \label{MH}
      \end{figure}
Fig.~\ref{MH} shows the $M$-$H$ curves, measured at different temperatures. At 200\,K, the magnetization is linear at higher magnetic field, along with slightly non-linear behavior and hysteresis at lower field values as shown in the inset. The similar behavior  is observed qualitatively for magnetization isotherms measured between 300  and 75 \,K. Below 75 \,K, coercivity varies significantly with a narrow hysteresis loop of the $M$-$H$ isotherms. The reduction in loop width coincides with the onset of magnetic transition in ZFC-FC measurements.
Below 30 \,K, the magnetization in $M$-$H$ measurements  shows a non-linear behavior at high fields. At 2 \,K, the $M$-$H$ isotherm  shows a completely different trend qualitatively as well as quantitatively.  A nearly saturation magnetization along with smaller coercivity is observed below 10\,K. A large magnetization value of nearly 4\,${\mu}_\mathrm{B}$/f.u. is attained at 5\,T magnetic field. This can be attributed mainly to the large magnetic moment associated with the Dy$^{3+}$ ion ($J$=15/2). At 2\,K, magnetization studies on single crystals of DyFeO$_{3}$ revealed that the magnetic moment along $a$ and $b$ crystallographic directions attain saturation values of 4.32\,${\mu}_{B}$/f.u. and 8.59${\mu}_{B}$/f.u. respectively, while along $c$-direction it shows a linear trend and attains a smaller value of 0.71\,${\mu}_{B}$/f.u. at 5\,T\cite{Zhao2014}. In our polycrystalline NdFeO$_{3}$ sample, the magnetic moment reaches a value of 1.28\,${\mu}_{B}$/f.u. in a field of 5\,T at 2\,K.
Thus considering of 50${\%}$ substitution by Nd atoms and polycrystalline nature of our sample, an approximate value of 2.91\,${\mu}_{B}$/f.u. for magnetization of NDFO is expected at 5\,T and 2\,K. This expected value of magnetic moments is much lower than the observed one (see Fig.~\ref{MH}), indicating a greater field induced polarization in NDFO. Below 10\,K, Dy$^{3+}$- Nd$^{3+}$ and  Nd$^{3+}$- Fe$^{3+}$ exchange interactions play crucial role in observation of higher value of magnetization induced by the rare-earth ordering/polarization\cite{Ankita2019}.  The hysteresis loop parameters viz. coercivity and retentivity at 2\,K  suggest a possible ordering of the rare-earth ions.\par
Thus, the $M$-$T$ (ZFC-FC) measurements and $M$-$H$ isotherms indicate the presence of a gradual spin reorientation below 75\,K. The FC measurements also rule out the possibility of a Morin transition. These DC magnetization measurements infer a possibility of rare-earth  (Dy$^{3+}$/Nd$^{3+}$) ordering at lower temperatures($<$10\,K).\par Magnetization measurements can not provide a conclusive evidence regarding the spin reorientation as well as rare-earth ordering. Thus, neutron diffraction measurements were carried out to understand the low temperature magnetic transitions in NDFO. 
%


    

\subsubsection{Neutron diffraction}
\label{neutron_diff}
  \begin{figure}[h!] \center
       \begin{picture}(250,320)
        \put(-12,-25){\includegraphics[height=370\unitlength,width=260\unitlength]{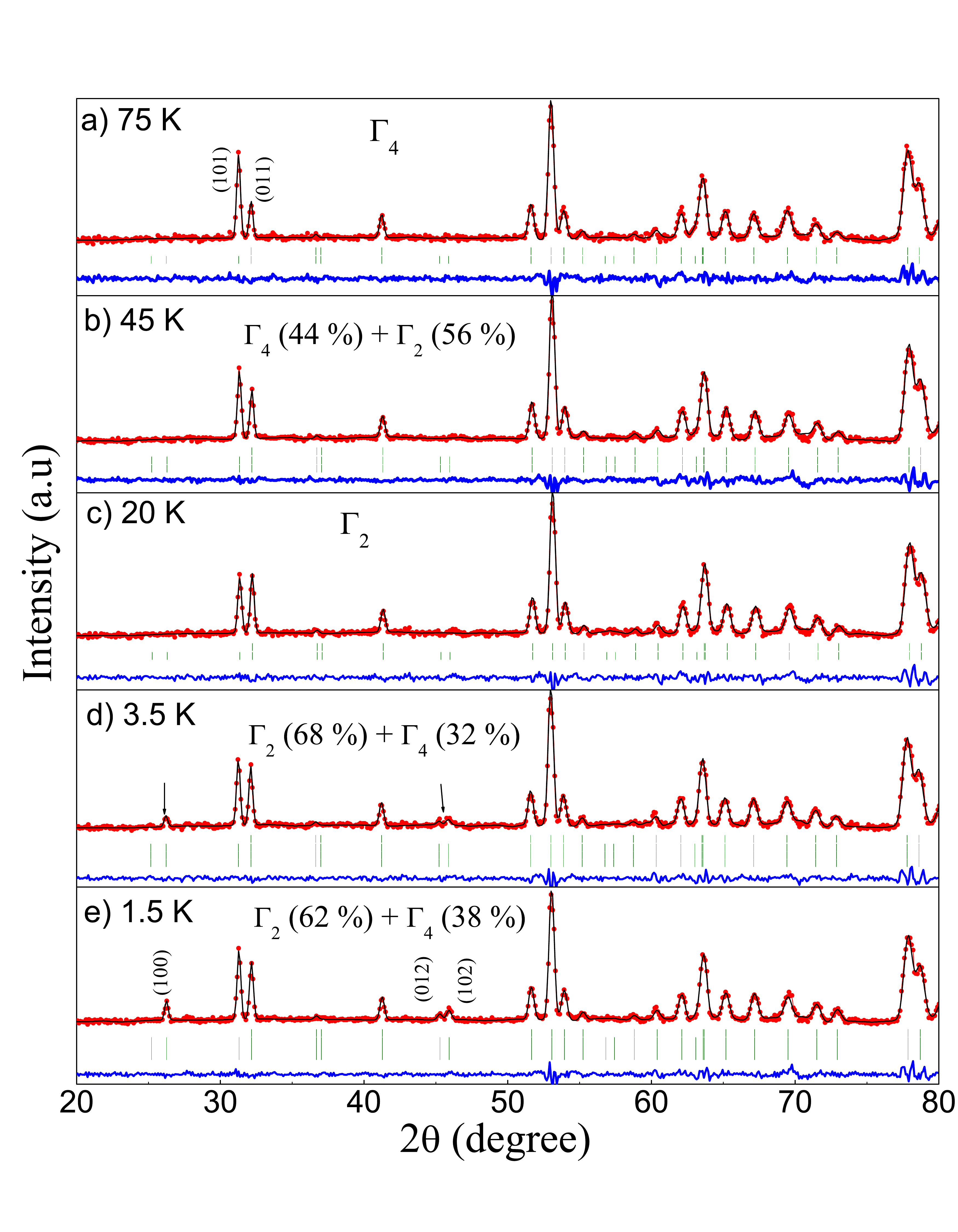}}
       \end{picture}
        \caption{(color online) Evolution of the neutron diffraction data along with refinement as a function of temperature.}
        \label{ND_temp}
      \end{figure}

  \begin{figure}[b!] \center
       \begin{picture}(260,170)
        \put(-15,-10){\includegraphics[width=280\unitlength,]{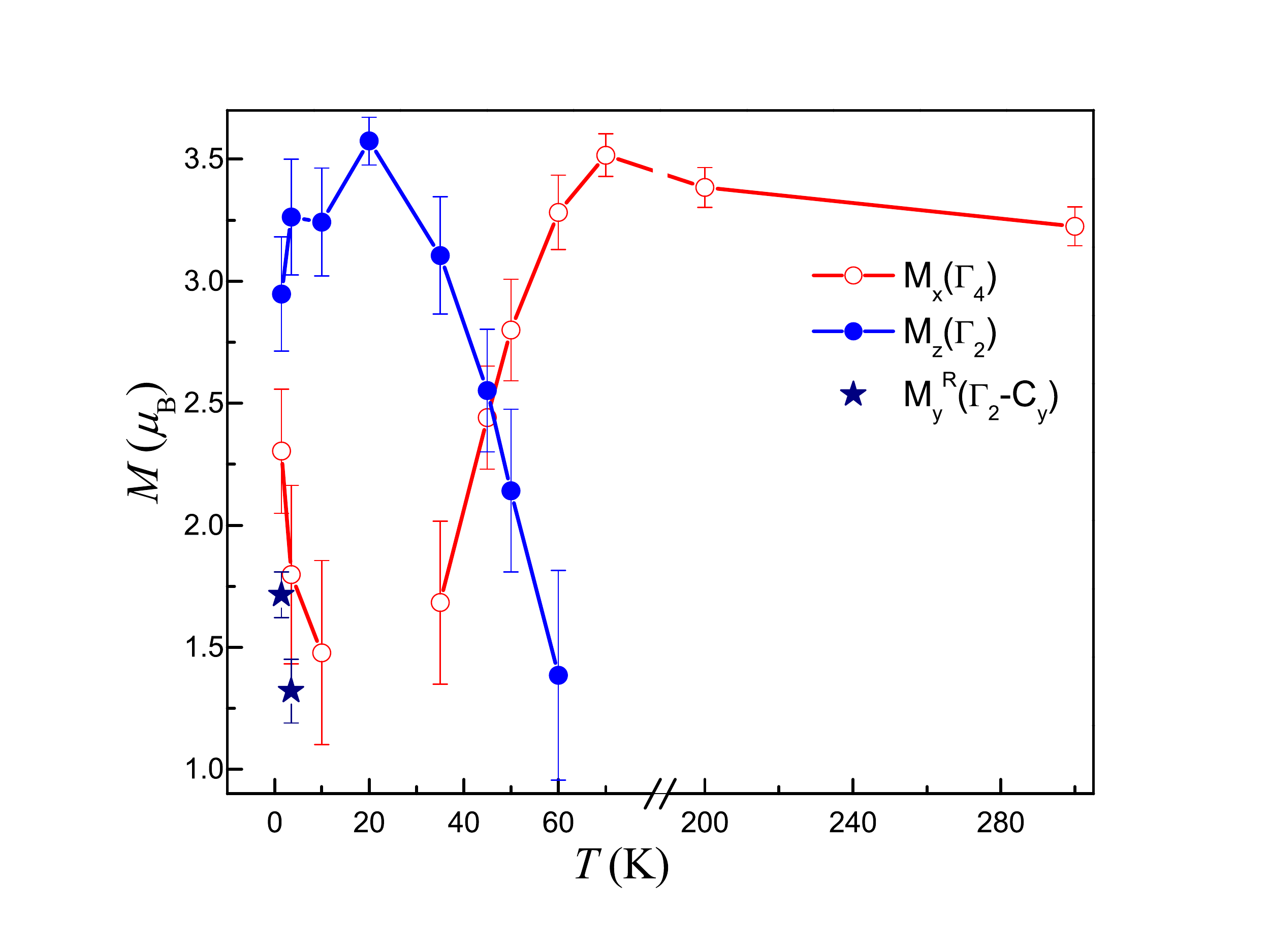}}
       \end{picture}
    \caption{(color online) Temperature variation of magnetic moments of Fe$^{3+}$ and Nd$^{3+}$/Dy$^{3+}$ spins from 1.5\,K to 300\,K for the various representations.}
        \label{mag_moments}
      \end{figure}

In this section, a systematic evolution of the NDFO magnetic structure, based on neutron diffraction data, is discussed. To obtain the detailed configurations of Fe$^{3+}$ and $R^{3+}$(Nd$^{3+}$/Dy$^{3+}$) magnetic moments in the unit cell, the magnetic structure has been solved from refinement of the powder neutron diffraction patterns for temperatures between 300 and 1.5\,K. The Fe atom occupies the $4b$ Wycoff position, while the $R$ atoms occupy the $4c$ sites.
 The magnetic ordering vector remains $k$=(0,0,0) in the entire temperature range for both Fe and $R$ atoms.
Eight irreducible representations, ${\Gamma}_{1}$ to ${\Gamma}_{8}$, exist from the symmetry analysis by Bertaut {\textit{et al.}}\cite{bertaut1963magnetism}.  Four out of these eight representations correspond to zero coefficients for the Fe site. Thus only four irreducible representations ${\Gamma}_{1}$ to ${\Gamma}_{4}$ can be considered which correspond to the Shubnikov magnetic space groups, ${\Gamma}_{1}$ ($Pbnm$), ${\Gamma}_{2}$ ($Pbn'm'$), ${\Gamma}_{3}$ ($Pb'nm'$), and ${\Gamma}_{4}$ ($Pb'n'm$). Using Bertraut's notation\cite{bertaut1963magnetism}, ${\Gamma}_{1},{\Gamma}_{2},{\Gamma}_{3}$, and ${\Gamma}_{4}$  representations can be written in a simplified manner as $G_y$, $F_x$$G_z$, $C_x$$F_y$, and $G_x$$F_z$ respectively corresponding to magnetic ordering of the cartesian components of Fe$^{3+}$ spins in the unit cell. Symbols $G$ and $C$ represent type of antiferromagnetic ordering and $F$ represents ferromagnetic component due to canting of antiferromagneticaly ordered Fe$^{3+}$ spins. Subscripts to the symbol represent the directions of Fe$^{3+}$ spins.\par 
Fig.~\ref{ND_temp} shows neutron diffraction patterns between 75 and 1.5\,K.
At 300\,K, the (101) and (011) magnetic peaks associated with G-type antiferromagnetic ordering of the Fe$^{3+}$ spins are observed near 2$\theta$= 32$^{\circ}$(data not shown). A ratio of 1/3 between the intensities of (101) and (011) magnetic peaks indicates that the magnetic structure belongs to the ${\Gamma}_{4}$ representation with a $G_{x}$ ordering of the Fe$^{3+}$ spins at 300\,K which is confirmed by refinement of the data. 
The ratio and intensities of (101) and (011) reflections do not vary down to 75 \,K.  The refined diffraction pattern at 75\,K is shown as Fig.~\ref{ND_temp} a). Below 75\,K,  the ratio ($I_\mathrm{(101)}/I_\mathrm{(011)}$) systematically increases with decreasing the temperature, indicating a change in the magnetic moment direction without modification of G-type magnetic structure i.e. spin reorientation. As shown in Fig.~\ref{ND_temp} a), b), and c), this variation in the ratio of peak intensities persists down to 20\,K, wherein the intensities of both the peaks have become almost equal ($I_\mathrm{(101)}/I_\mathrm{(011)}$=1).  At 20\,K, the presence of magnetic ordering represented by ${\Gamma}_{2}$ irreducible representation ($F_x$, $G_z$) only is evident due to equal intensities of the (011) and (101) peaks.  Below 10 \,K the intensity of the (101) peak again increases w.r.t the (011) peak  which suggests a second spin reorientation/reoccurrence of magnetic phase (${\Gamma}_{2}\rightarrow{\Gamma}_{4}+{\Gamma}_{2}$). 
At the lowest temperatures viz. 3.5 and 1.5\,K (Fig.~\ref{ND_temp} d) and e)),  three additional peaks develop at $2\theta$=25$^\circ$ and 45$^\circ$, which marks the rare-earth magnetic ordering.\par 
In NDFO, from 300\,K till 75\,K, the magnetic structure belongs to ${\Gamma}_{4}$ representation, wherein the Fe spins are arranged  in $G_{x}$ type  antiferromagnetic structure with a weak ferromagnetic component  ($F_{z}$) along crystallographic $c$-direction. From the diffraction pattern, we do not find any peak corresponding to the ferromagnetic $F_{z}$ component i.e. the (002) peak near $2\theta$=40$^\circ$. The peak might be undetected due to small values of canting angle leading to a small ferromagentic moment of less than 0.1 ${\mu}_{B}$/f.u. which is not seen from our powder diffraction experiments. However the spin configuration matches with  the orthoferrites.\par
 %
\begin{figure*}[t!] 
\includegraphics[width=\textwidth]{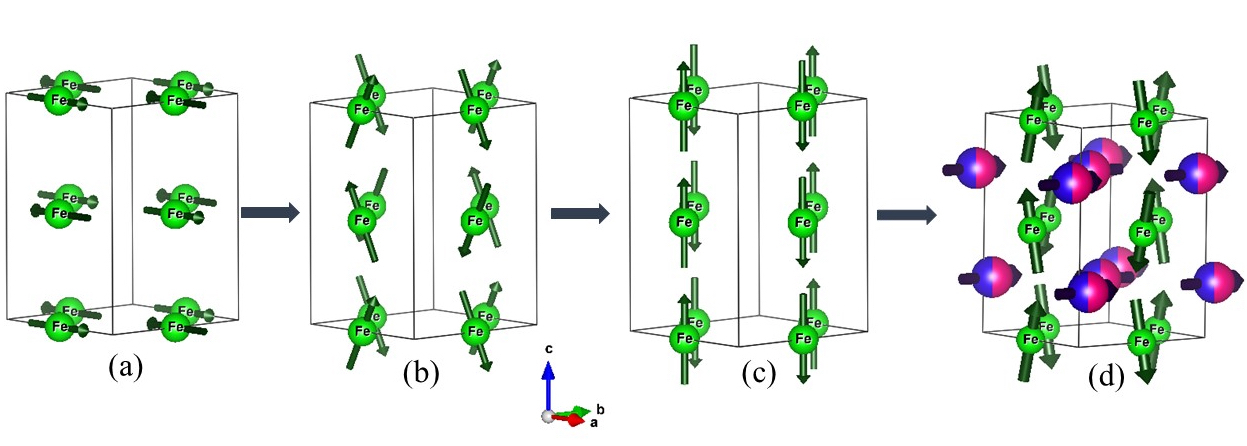}
        \caption{(color online)  Schematic representation of NDFO magnetic structure at (a) 300 \,K $\Gamma_4$ (b) reorientation region at 45 \,K: $\Gamma_{42}$ (c) at 20 \,K : $\Gamma_2$ and (d) 1.5 \,K second reorientation : $\Gamma_{24}$ and ordering of Nd/Dy as $C_{y}$.}
        \label{ND_mag_struct}
\end{figure*}
Below 60\,K, the magnetic structure is best refined with the mixture of ${\Gamma}_{2}$ and ${\Gamma}_{4}$ representations, indicating the ongoing process of spin reorientation. As shown in Fig.~\ref{ND_temp} b), the magnetic structure of NDFO consists 44\,\%${\Gamma}_{4}$ and 56\,\%${\Gamma}_{2}$ phases at 45\,K. The refinements down to 20 K clearly indicate that a gradual ${\Gamma}_{4}$${\rightarrow}$${\Gamma}_{2}$ type spin reorientation takes place in NDFO which is similar to the the usual second order spin reorientation observed in NdFeO$_{3}$\cite{Yuan2013}. This is in contrast with the ${\Gamma}_{4}$${\rightarrow}$${\Gamma}_{1}$ type abrupt spin reorientation observed in DyFeO$_{3}$\cite{yamaguchi1974,Wang2016}. 
As indicated by the equal ratio of $I_\mathrm{(101)}$ and $I_\mathrm{(011)}$ at 20\,K, our analysis confirmed that the magnetic structure belongs entirely to ${\Gamma}_{2}$ representation. Inclusion of ${\Gamma}_{4}$ leads to higher values of ${\chi}^{2}$. At 10 \,K and below, high temperature magnetic phase (${\Gamma}_{4}$) again starts to reappear and the magnetic structure belongs to a mixture of ${\Gamma}_{2}$+${\Gamma}_{4}$ representations. Interestingly, below 10\,K the volume fraction of ${\Gamma}_{4}$ increases grudually, while that of ${\Gamma}_{2}$ reduces, a trend which is reverse to the observed spin reorientation between 60-20\,K. As shown in Fig.~\ref{ND_temp} d)\,$\&$\,e), this trend is visible clearly from the variation in intensities of (101) and (011) peaks. At lowest measured temperature (1.5\,K), 62\,\% ${\Gamma}_{2}$  and 38\,\% ${\Gamma}_{4}$ phases constitute the antiferromagnetic order of Fe$^{3+}$ spins in NDFO.
\par
At 3.5\,K, the (100) magnetic peak at 25$^{\circ}$ and (012) and (102) magnetic peaks around $2\theta$=45$^\circ$, pertaining to rare-earth ordering, arise due to the $c_{y}^R$ type arrangement of $R^{3+}$ (Dy$^{3+}$/Nd$^{3+}$) moments which also belongs to the ${\Gamma}_{2}$ representation.
In NDFO, the C-type ordering of Dy$^{3+}$/Nd$^{3+}$ moments is unusual since the Dy$^{3+}$-Dy$^{3+}$ interactions and strong single ion anisotropy of the Dy$^{3+}$ moments would result a ($g_{x}^R$, $a_{y}^R$) ordering\cite{Yamaguchi1973}. However in NDFO, the neutron diffraction patterns do not show a signature of this ordering  till 1.5\,K.
\par
The temperature variation of the magnetic moments for the Fe$^{3+}$ and $R^{3+}$ moments for different magnetic structures is shown in Fig.~\ref{mag_moments}. From 300\,K till 75 \,K, a small increase, in the magnetic moment associated with the ${\Gamma}_{4}$($G_{x}$) representation of Fe$^{3+}$ spins, is observed. Below 60 \,K with the onset of first spin reorientation, there is a systematic decrease in the magnetic moment along crystallographic $a$ direction ($M_x$), while correspondingly the magnetic moment along $c$ direction ($M_z$) shows an increase. $M_z$ attains maximum value at 20\,K conforming the presence of magnetic structure represented by ${\Gamma}_{2}$ only.  At 10\,K and below, the $M_z$ magnetic moment starts to decrease, while the $M_{x}$ again starts to increase. The values of total magnetic moment of Fe$^{3+}$ is nearly 3.7\,${\mu}_\mathrm{B}$, which is lower than the theoretical (ionic) value 5\,${\mu}_\mathrm{B}$. Such a reduction in magnetic moment can be due to the effects of covalency and polycrystalline nature of our sample.
On the other hand, the magnetic moment of Nd$^{3+}$/Dy$^{3+}$ is nearly 1.8 ${\mu}_\mathrm{B}$ at 1.5\,K.  
However, this value of magnetic moment (1.8 ${\mu}_\mathrm{B}$)  of Nd$^{3+}$/Dy$^{3+}$ is twice the value of 0.9 ${\mu}_\mathrm{B}$ which was obtained from neutron diffraction measurements on the single crystals of NdFeO$_{3}$\cite{bartolome1997single}. This indicates the greater polarization of the rare-earth sub-lattice due to $R^{3+}$-Fe$^{3+}$ exchange interaction, which also causes the alignment of highly anisotropic Dy$^{3+}$ magnetic moments. As discussed in sections~\ref{theoretical}\&\ref{discuss}, the enhancement of the  polarization and magnetic ordering of rare-earth moments are explained by calculating the strength of various exchange interaction using DFT. The schematic representation of  Fe$^{3+}$ and  rare-earth (Nd$^{3+}$/Dy$^{3+}$)  magnetic structure along with temperature dependent successive spin reorientation is depicted in the Fig.~\ref{ND_mag_struct}.
\subsection{Heat Capacity}
\label{Spec_heat}
\begin{figure}[h!] 
\center
       \begin{picture}(500,480)
        \put(0,0){\includegraphics[height=500\unitlength,width=250\unitlength]{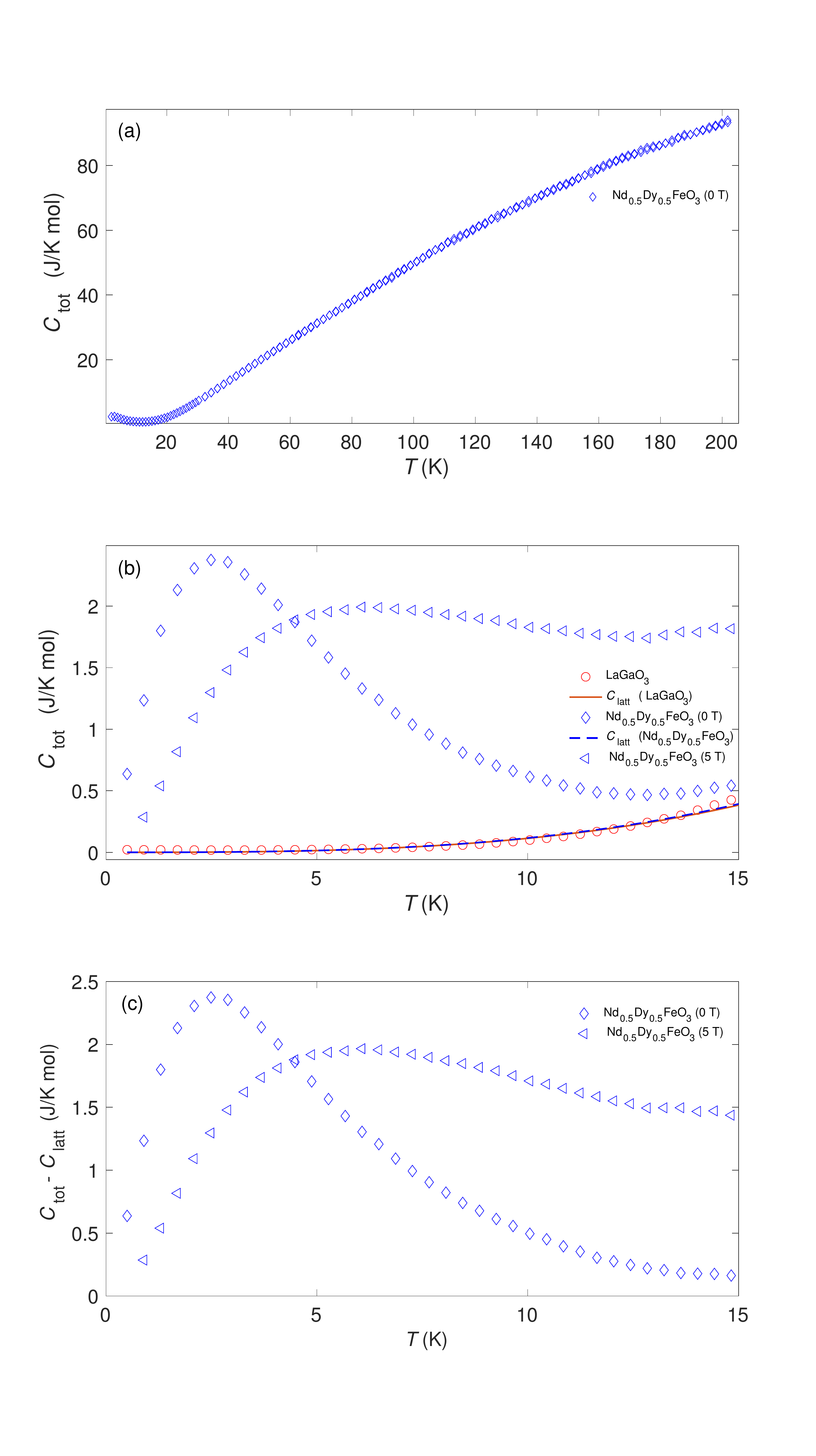}}
       \end{picture}
        \caption{(color online) (a) The molar heat capacity of NDFO measured at 0\,T over 2-220\,K. (b) The molar heat capacity of NDFO and non-magnetic analog LaGaO$_{3}$ measured at 0\,T over 0.4-15\,K. Solid red line shows calculated lattice contribution to the molar heat capacity for LaGaO$_{3}$. Dashed blue line: renormalized lattice contribution for NDFO. Open triangles show molar heat capacity of NDFO at 5\,T over 0.4-15\,K. (c)  Molar heat capacity for NDFO at 0 and 5\,T after subtracting the lattice contribution.}
        \label{Specific_heat}
      \end{figure}

The molar heat capacity data of NDFO are shown in Fig.~\ref{Specific_heat}(a) for the temperature range from 2 to 200\,K. The heat capacity data could not identify any distinct signature associated with the spin reorientations of Fe$^{3+}$ moments within the limit of measurement resolution. The heat capacity measurements were extended down to 0.4\,K to investigate the possible rare-earth (Nd/Dy) ordering. The molar heat capacity data over 0.4-15\,K under 0 and 5\,T are shown in Fig.~\ref{Specific_heat}(b).\par To estimate the lattice contribution in the heat capacity, an isomorphous non-magnetic compound LaGaO$_{3}$ was synthesized and its heat capacity was measured over 0.4-15\,K. The lattice contribution for LaGaO$_{3}$ follows Debye function with  effective Debye temperatures of 390 and 495\,K for (La\,\&\,Ga) and O atoms, respectively\cite{Kumar2016}. The lattice contribution for the heat capacity of NDFO has been estimated using the Debye function with effective Debye temperature of 495\,K for oxygen atoms and  384.8\,K (renormalized with the mass) for (Nd/Dy\,\& Fe).\par
The molar heat capacity data of NDFO after removing the lattice contribution are shown in Fig.~\ref{Specific_heat}(c). Below 10 \,K,  a broad peak in the heat capacity is observed at $\sim$2.2\,K which shifts to higher temperature under an applied magnetic field of 5\,T.  The observed broad peak in NDFO could originate from Schottky anomaly due to crystal-field splitting of $4f$ electronic states in Nd$^{3+}$ and Dy$^{3+}$ ions. We could not satisfactorily fit the observed peak and its field dependence by considering only the Schottky term associated with crystal field splitting of Nd$^{3+}$ and Dy$^{3+}$ ions. This shows that the magnetic ordering of the rare-earth also contribute to the observed broad peak of low temperature ($<$10\,K) heat capacity data.  In NDFO, presence of two Kramer's ions can cause additional complexity due to entirely different strengths of Nd$^{3+}$-Fe$^{3+}$ and Dy$^{3+}$-Fe$^{3+}$ exchange interactions which lift the degeneracy of both the doublets. In addition, the $R$-$R$ exchange interactions also split the doublet in a more complex manner.\par
 The ${\lambda}$-shaped anomaly associated with second order phase transition of rare-earth ordering is not observed till 0.4\,K. The ${\lambda}$-shaped anomaly is seen prominently in DyFeO$_{3}$ at 4.2\,K\cite{Zhao2014}. On the other hand, NdFeO$_{3}$ shows a broad Schottky peak along with a less noticeable ${\lambda}$ anomaly superposed on the Schottky anomaly at 1.05 \,K\cite{Bartolome1994}. Thus, as per our measurements, a distinct discontinuity (second order transition) in heat capacity due to independent ordering of the rare-earth is completely absent in NDFO. Thus the ordering of rare-earth in NDFO is driven by the effects of molecular field due to Fe$^{3+}$ spins, rather than independent rare-earth ordering. 

\section{DFT results}
\label{theoretical}
\begin{figure}[b!] \center
       \begin{picture}(250,130)
        \put(0,-5){\includegraphics[width=270\unitlength]{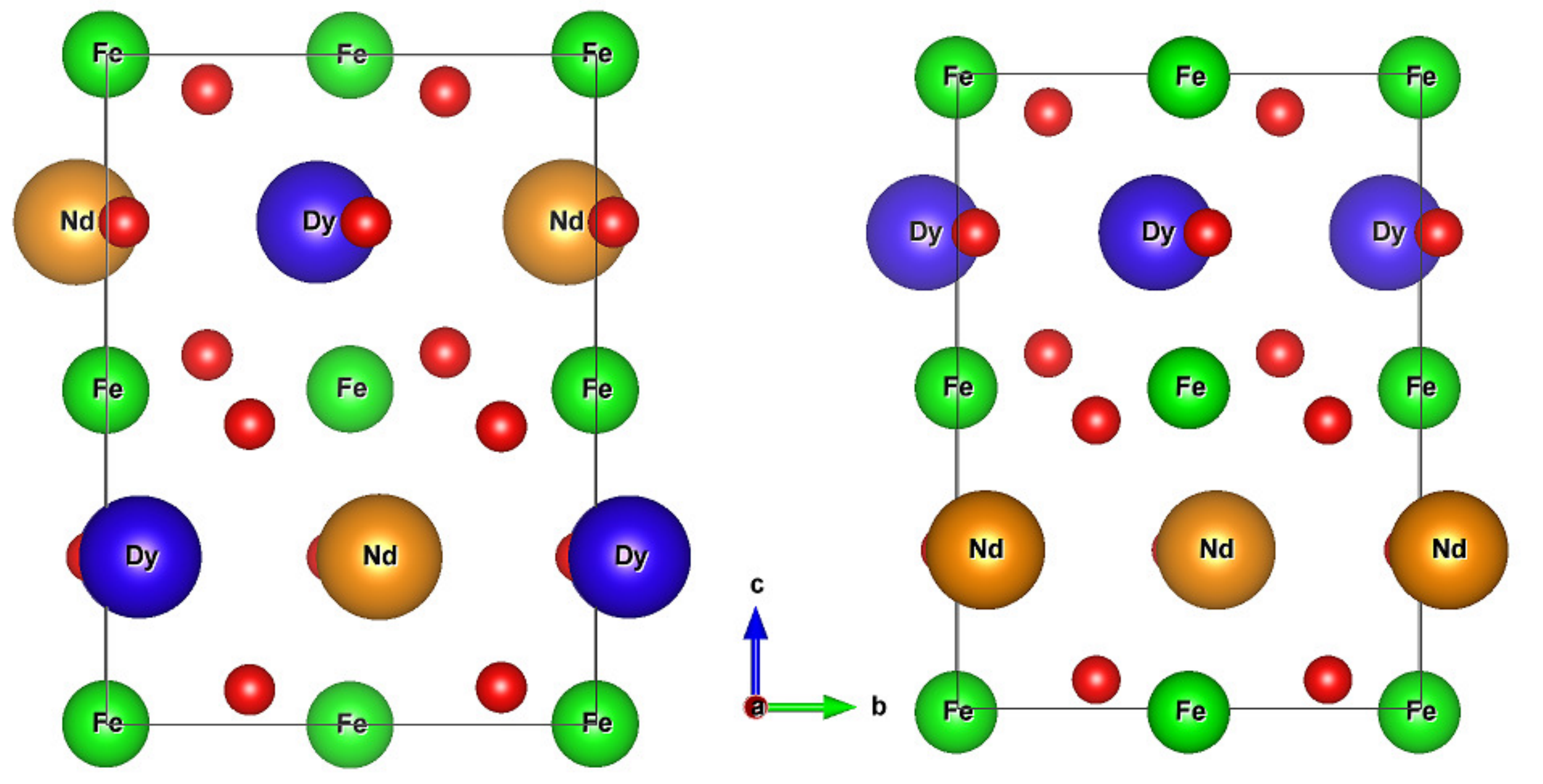}}
       \end{picture}
        \caption{(color online) Unit cells of NDFO displaying two possible arrangements of Nd and Dy ions considered in our calculations: (a) alternate (b) layered.}
        \label{unit_cells}
      \end{figure}

To understand the complex interplay between the Nd$^{3+}$/Dy$^{3+}$ and Fe$^{3+}$ magnetic moments, the ground state magnetic order is evaluated using density functional theory. Since the Nd and Dy atoms occupy the same crystallographic site, their occupancies are random. However for computational purposes,
we have considered two possible arrangements of the Nd and Dy: alternate (or 111) and layered (or 001). In the alternate (111) arrangement, the Nd and Dy atoms are placed adjacent to each other. Thus each Nd atom has six Dy atoms as nearest neighbours and vice versa. In the layered (001) arrangement, the planes of Nd and Dy atoms are alternately stacked along the $c$ direction\cite{Ankita2017}. Fig.~\ref{unit_cells} depicts these two possible arrangements; (a) alternate (111) and (b) layered (001). These two different arrangements give us scope to probe the nature of Nd$^{3+}$-Nd$^{3+}$, Dy$^{3+}$-Dy$^{3+}$ and Nd$^{3+}$-Dy$^{3+}$ exchange interactions. 
Structural relaxation of the orthorhombic unit cell with the experimental lattice parameters obtained at 1.5 \,K has been performed for two possible arrangements of Nd and Dy atoms discussed above. The structure was relaxed considering G-type magnetic ordering of the Fe$^{3+}$ magnetic moments. The $4f$ electrons of Nd and Dy are treated as core electrons during relaxation.
To obtain the ground state magnetic ordering of Fe$^{3+}$ and Nd$^{3+}$/Dy$^{3+}$ sub-lattices, the electronic self-consistent calculations are performed considering the effects of Coulomb correlation $U$ for the Fe $3d$ states and Nd/Dy $4f$ states. The values of Hubbard parameters used (Coulomb correlation $U$ and Hund's exchange $J$) are such that $U_\mathrm{eff}$ = $U$-$J$ = 6.0\,eV for Nd and Dy, while for Fe, $U_\mathrm{eff}$ = 3.0\,eV which are also used in previous literature\cite{Chen2012}. The self-consistent calculations are performed till an energy difference of  10$^{-6}$\,eV between successive iterations is achieved.

\begin{figure*}[tbh]
\begin{minipage}{0.5\linewidth}
\centering
\includegraphics[width=9.0cm]{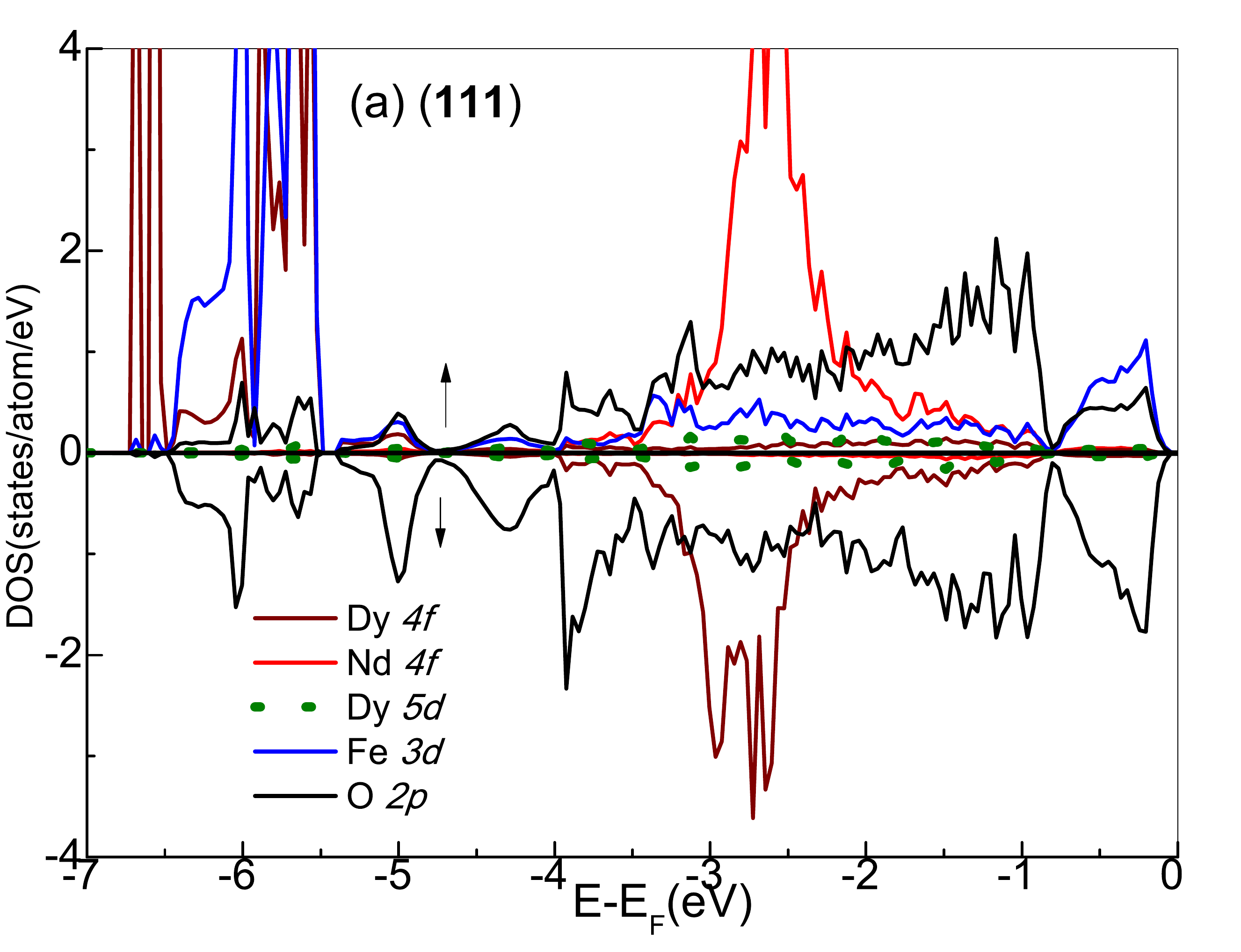}
\end{minipage}%
\begin{minipage}{0.5\linewidth}
\hspace*{0.1cm}
\centering
\includegraphics[width=9.0cm]{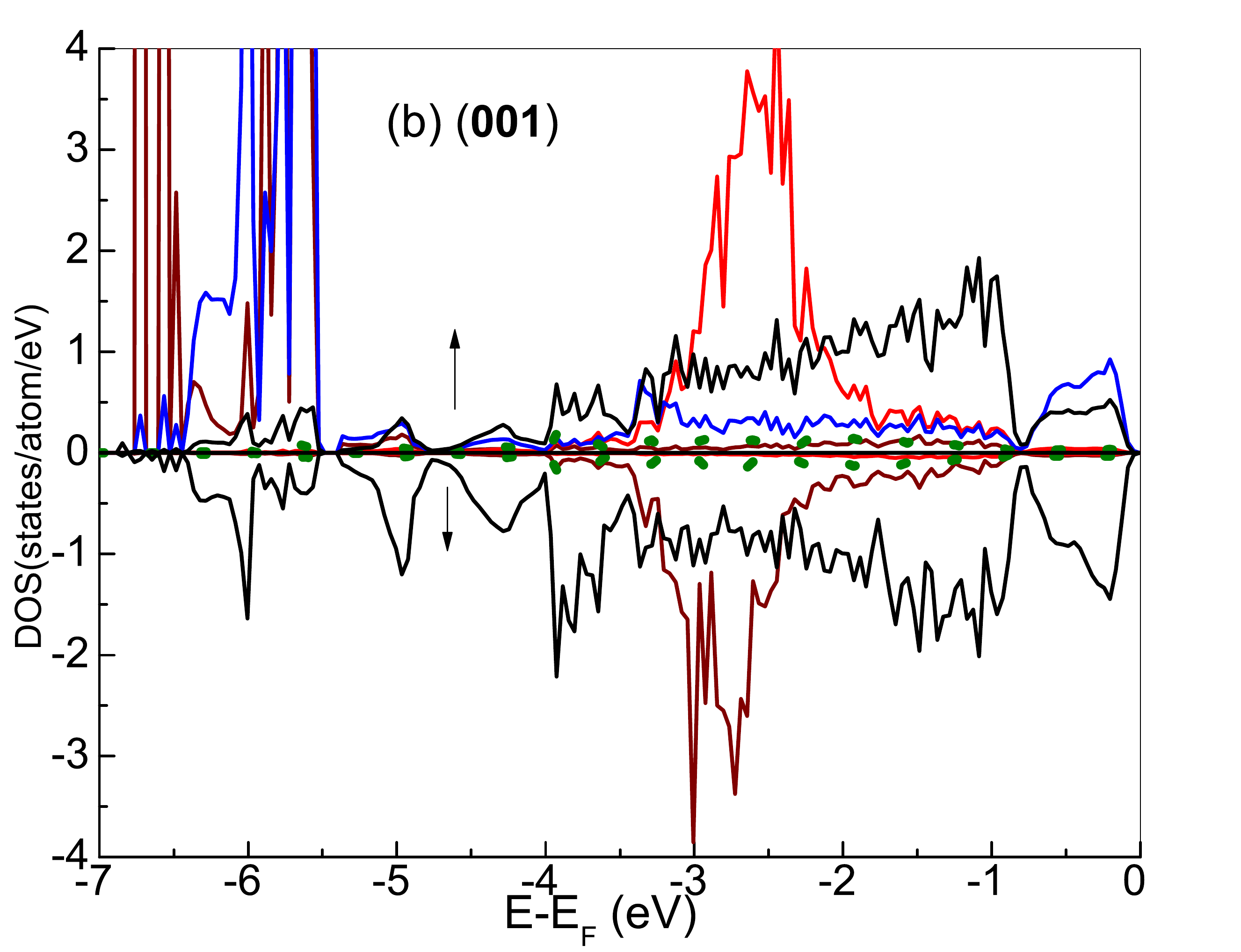}
\end{minipage}
\caption{(color online) Spin resolved density of states (DOS) of NDFO for the (a) alternate(111) (b) layered(001) arrangements of Nd and Dy corresponding to the $C$ type ordering of rare-earth and $G$ type ordering of Fe moments. The ${\uparrow}$ and ${\downarrow}$ correspond to spin up and down regions respectively.} 
\label{DOS}
\end{figure*}

The ground state magnetic structure was determined by comparing total energies of various arrangements of Fe$^{3+}$ and Nd$^{3+}$/Dy$^{3+}$ magnetic moments. As the magnetic ordering of Fe$^{3+}$ sub-lattice occurs at a much higher temperature compared to that of the rare-earth ordering, we have performed calculations of collinear A, C and G-type antiferromagnetic configuration along with ferromagnetic one in the absence of magnetic ordering of the rare-earth ions (i.e. we considered $4f$ electrons as core electrons for this purpose). The lowest energy in this case is obtained for G-type magnetic ordering which is the preferred ordering of Fe$^{3+}$ spins in all orthoferrites below N\'eel temperature and prior to spin reorientation transition.

To probe the rare-earth ordering, the Fe$^{3+}$ magnetic moments are fixed as G-type in the calculations. In case of parent compounds NdFeO$_{3}$ and DyFeO$_{3}$, the rare-earth moments order in C- and G-type AFM respectively\cite{Przeniosto1995}. Therefore, in NDFO, the G-type ordering of rare-earth moments is as likely to occur as that of C-type. From our neutron diffraction measurements discussed above we have observed C-type ordering of rare-earth ordering. To support our experimental findings, we have performed the self consistent total energy calculations within GGA+$U$ approximation for both  C-type (experimentally observed one) and G-type order of rare-earth moments keeping Fe moments in G-type order. We observe from our calculations that for both the structural arrangements (111 and 001) of Nd and Dy, C-type magnetic order emerges  as the lowest energy which is consistent with our neutron diffraction results. The relative energies are listed in Table~\ref{table2} wherein the C-type is set to 0\,meV. As seen, for the alternate arrangement of Nd/Dy atoms, the G-type ordering has a higher energy as compared to the layered arrangement. Thus it is established from our experimental observations as well as our DFT calculations that the rare-earth ions order in C-type AFM in NDFO.

\begin{table}
\caption{Relative energies (in meV) for two main antiferromagnetic orders of $R^{3+}$ ions within GGA+$U$ ($U_\mathrm{eff}$=6\,eV at Nd/Dy and 3\,eV at Fe sites) for the two different Nd/Dy arrangements.  }
\begin{center}
\begin{tabular}{p{4.0cm}c c c c c}\hline\hline 
Magnetic structure & alternate &  layered &\\ \hline  
C-type & 0 & 0   \\ 
G-type & +55 & +30  \\ 
\hline\hline
\end{tabular}
\end{center}
\label{table2}
\end{table}
Further to look into the electronic structure of our compound, we calculated the electronic density of states (DOS) for the ground state magnetic order observed in our calculations (i.e. rare-earth ions order in C-type and Fe$^{3+}$ ions order in G-type antiferromagnetic configuration) for both (111) and (001) Nd/Dy arrangements.
Irrespective of the cationic arrangements, we observe that the electronic structure of Fe remains unaffected. 
 Figs.~\ref{DOS}(a) and (b) show the spin resolved density of states for each atom of NDFO for (111) and (001) arrangements, respectively, below the Fermi energy. In both arrangements, the energy difference between the highest occupied and lowest unoccupied states (not shown in figure) correspond to a band gap of ${\sim}$ 2\,eV. The Fe $3d$ states are strongly hybridized with the O $2p$ states and show same spectral character in both the cationic arrangements.  
Below $E_\mathrm{F}$, the Nd/Dy $4f$ states show a broader peak, centered at -2.5\,eV. The $4f$ states show greater hybridization with the O $2p$ states in this region. This region also contains contribution from the Dy $5d$ states which mediate the Dy-O-Nd hybridization. Sharp peaks corresponding to the Dy $4f$ states occur near -6\,eV, which are highly localized. 
%
%

\begin{figure}[h!] \center
       \begin{picture}(500,180)
        \put(0,-5){\includegraphics[width=220\unitlength]{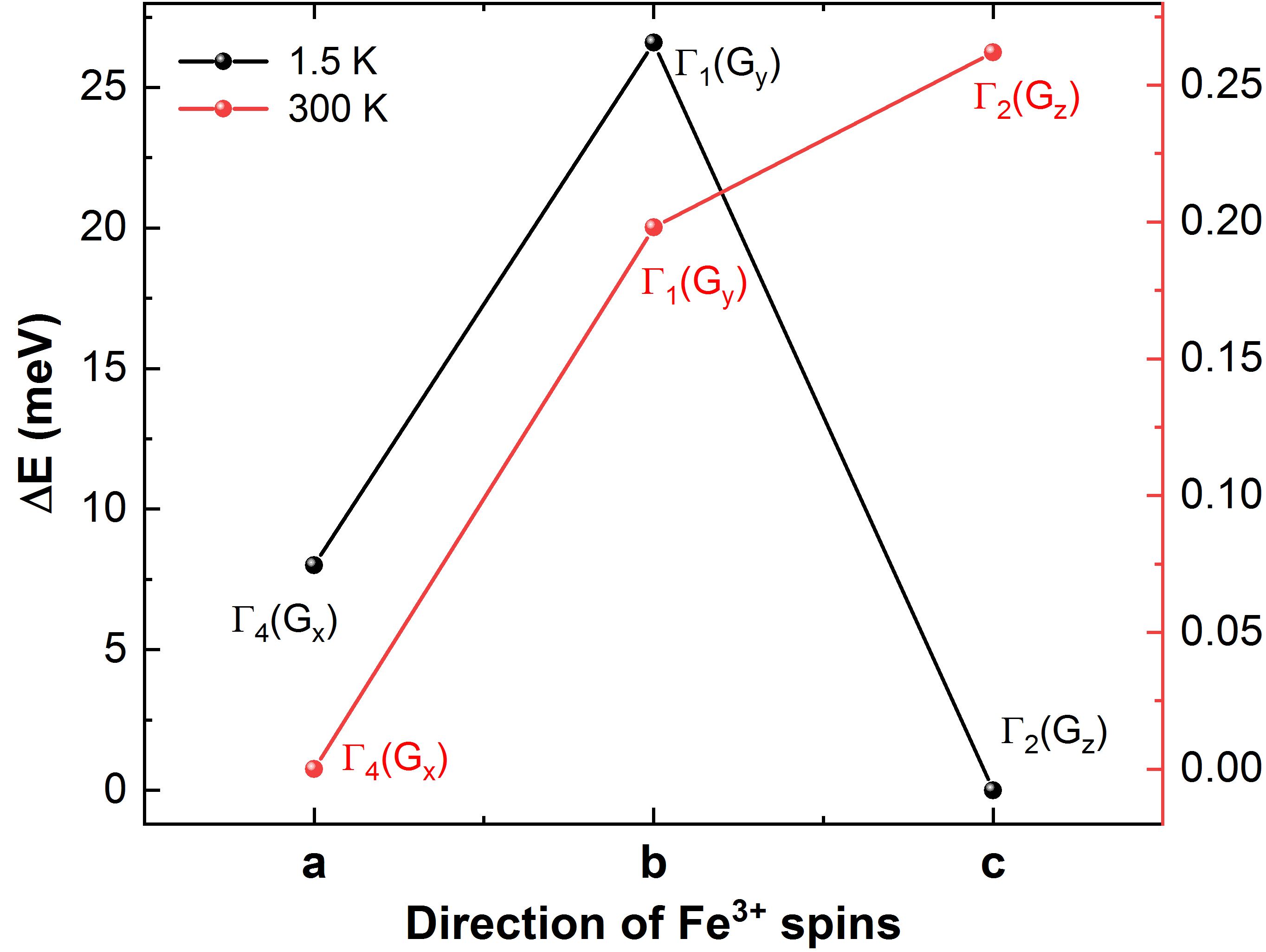}}
       \end{picture}
        \caption{(color online) Relative energies (in meV per unit cell) calculated from non-collinear calculations within GGA+$U$+SO for Fe$^{3+}$ spins along the three crystallographic directions for 300K (red) and 1.5K (black) crystal structures.}
        \label{energy_comparision}
      \end{figure}

Most difficult part to probe is the spin-reorientation behavior of Fe$^{3+}$ moments. 
To understand the role of anisotropy and the ${\Gamma}_{4}$${\rightarrow}$${\Gamma}_{2}$ spin reorientation in NDFO, non-collinear calculations, with spin-orbit coupling within GGA+$U$+SO approximation, are performed for relaxed structural parameters corresponding to 300 \,K and  1.5 \,K. These calculations are performed only for the case of (111) arrangement of the Nd/Dy atoms as this arrangement is found to be lower in energy than the (001) arrangement.  
As $R^{3+}$-Fe$^{3+}$ interactions are nearly an order of magnitude smaller than the Fe$^{3+}$-Fe$^{3+}$ exchange interactions, for 300 \,K the $4f$ electrons of Nd and Dy are considered as core electrons and thus non-magnetic. In such a scenario, we performed calculations
with Fe$^{3+}$ moments pointing along crystallographic $a$, $b$ and $c$-axes in G-type antiferromagnetic arrangement. The trend in relative energies from our non-collinear calculations are shown in Fig.~\ref{energy_comparision}. It is clearly seen that in case of 300K structure, the magnetic order with Fe moments pointing along  crystallographic $a$ direction (${\Gamma}_{4}$($G_{x}$)),  is the preferred one which is in agreement with our neutron data.
For $T$ = 1.5 \,K, the $4f$ electrons of Nd and Dy are considered as valence electrons with magnetic moment fixed according to $c_{y}^R$ component of ${\Gamma}_{2}$ representation. The calculations were performed for the three different directions of Fe magnetic moments corresponding to $c$(${\Gamma}_{2}$), $b$$({\Gamma}_{1}$) and $a$(${\Gamma}_{4}$) directions. 
From Fig. ~\ref{energy_comparision} we see that the $c$(${\Gamma}_{2}$) direction of Fe$^{3+}$ has lowest energy indicating it to be easy axis of Fe$^{3+}$ spins, consistent with experimental results and symmetry analysis. On the other hand, the $b$ direction which is easy axis of Fe$^{3+}$ spins in DyFeO$_{3}$, has highest energy. Interestingly, one also observes from Fig.~\ref{energy_comparision} that in the low temperature phase the $c$(${\Gamma}_{2}$) and $a$(${\Gamma}_{4}$) configurations are energetically closer to each other than $b$$({\Gamma}_{1}$) whereas in high temperature phase $c$(${\Gamma}_{2}$) and $a$(${\Gamma}_{4}$) configurations are well separated in energies. This observation can be directly correlated with our neutron measurements where it is seen that at high temperatures there exists a pure $a$(${\Gamma}_{4}$) phase whereas at lowest temperatures, a mixed phase of $c$(${\Gamma}_{2}$) and $a$(${\Gamma}_{4}$) emerges. This implies significant role of Nd/Dy 4f states in the reappearance of $a$(${\Gamma}_{4}$) phase at lowest temperatures.  


In the orthoferrites, below the spin-reorientation temperature, the $R^{3+}$-$R^{3+}$ exchange interactions compete with the $R^{3+}$-Fe$^{3+}$ exchange interactions. 
In the Nd-based isostructural compounds where the B-site atom is non-magnetic, for instance, NdGaO$_{3}$ and NdScO$_{3}$, the Nd$^{3+}$ moments order as G-type\cite{Luis1998, Plaza1997}. 
The highly anisotropic Nd$^{3+}$-Cr$^{3+}$ exchange interactions in NdCrO$_{3}$ though polarize the Nd$^{3+}$ moments, clearly affect and eventually suppress the Nd$^{3+}$ ordering\cite{Bartolome1994}. Similarly, in NdFeO$_{3}$, Nd$^{3+}$-Fe$^{3+}$ exchange interactions cause polarization of the Nd$^{3+}$ moments, which thereby order as $c_{y}^R$ shifting the Nd$^{3+}$ ordering to $T_\mathrm{N2}$${\sim}$1.05 \,K. On the other hand, in DyFeO$_{3}$, the Dy$^{3+}$ moments order independently as ($g_{x}^R$, $a_{y}^R$), indicating negligible polarization of Dy$^{3+}$ due to Fe$^{3+}$ moments. Thus in NDFO, it would be interesting to obtain an idea of the order of various exchange interactions as the Nd$^{3+}$-Fe$^{3+}$ anisotropic exchange interactions seem to play a much larger role compared to the large single ion anisotropy of Dy$^{3+}$ ion and the Dy$^{3+}$-Dy$^{3+}$ interactions. 
We have determined the strengths of the Nd$^{3+}$-Nd$^{3+}$, Dy$^{3+}$-Dy$^{3+}$ and Nd$^{3+}$-Dy$^{3+}$ exchange interactions within GGA+$U$ approximation ($U_\mathrm{eff}$=6\,eV). Additionally the strength of Nd$^{3+}$/Dy$^{3+}$-Fe$^{3+}$ exchange interactions are also determined. The exchange interaction strengths are evaluated by mapping the energy difference between the ferromagnetic and antiferromagnetic configurations to the Heisenberg spin Hamiltonian as per the method used by Weingart {\it et. al.}\cite{Spaldin2012}.\par

\begin{table}[]
\caption{In-plane ($ab$) and out of plane (along $c$) exchange interaction strength ($J$) between the $R^{3+}$ and Fe$^{3+}$ in NDFO. '+' sign corresponds to AFM, while `-' sign corresponds to FM interactions.}
\begin{center}
\begin{tabular}{p{4.0cm}c c c c c}\hline\hline
Exchange Interaction & $J_{ab}$ (meV)  & $J_{c}$ (meV)    && \\ \hline
Nd$^{3+}$-Nd$^{3+}$             & -0.028 & -0.40      & & \\ \hline
Dy$^{3+}$-Dy$^{3+}$             & -0.31 & -1.8     & & \\ \hline
Nd$^{3+}$-Dy$^{3+}$             & 0.40  & -0.31    & & \\ \hline
Nd$^{3+}$-Fe$^{3+}$             & -0.62 & -0.68    & & \\ \hline
Dy$^{3+}$-Fe$^{3+}$             & 0.015  & 0.020  &  &  \\ \hline
Fe$^{3+}$-Fe$^{3+}$             & 5.40  & 5.28    & & \\ \hline 
\hline\hline
\end{tabular}
\end{center}
\label{table3}
\end{table}
The calculations were performed on ``artificial unit cells" in which except for the selected Fe or Nd/Dy atoms, the rest of magnetic atoms were replaced by non-magnetic atoms. Thus the Fe atom is replaced by Al, while Nd and Dy are replaced by La atoms.  The Al$^{3+}$ and La$^{3+}$ ions are non-magnetic. The interaction strengths ($J$) are determined between the pair of atoms that are a) in $ab$ plane and b) out-of-plane along $c$ direction and are listed in Table~\ref{table3}. The Fe$^{3+}$-Fe$^{3+}$ interaction which is strongest, is anti-ferromagnetic in-plane and out-of-plane corresponding to the G-type magnetic ordering. The highly isotropic nature is seen from the small difference between $J_{ab}$ and $J_{c}$.   
Compared to Fe$^{3+}$-Fe$^{3+}$, rest of the interactions are nearly an order of magnitude smaller. The interaction strength between Nd$^{3+}$-Fe$^{3+}$ has the second highest value which determines the magnetic ordering of the $R^{3+}$ moments below the reorientation region. On the other hand, the Dy$^{3+}$-Fe$^{3+}$ exchange interaction is the weakest. This is in agreement with previous experimental studies which show that the Dy$^{3+}$-Fe$^{3+}$ exchange field in DyFeO$_{3}$ is ${\sim}$ 2 T\cite{Zvezdin1979}, while in NdFeO$_{3}$, the Nd$^{3+}$-Fe$^{3+}$ exchange field is nearly 6.6 T\cite{bartolome1997single}. 
The Nd$^{3+}$-Nd$^{3+}$ and Dy$^{3+}$-Dy$^{3+}$ exchange interactions are highly anisotropic. The magnitude of the later is considerably higher along the $c$ direction. This is expected since in parent compound DyFeO$_{3}$ the Dy$^{3+}$ moments order at nearly 4.5\,K. However in NDFO 50${\%}$ substitution by Nd tends to suppress the Dy$^{3+}$-Dy$^{3+}$ interactions.  
Most importantly, the nature of Nd$^{3+}$-Dy$^{3+}$ exchange interaction is clearly C-type with opposite signs for in-plane (AFM) and out of plane (FM) interactions. Moreover, the competing exchange interaction strengths of Nd$^{3+}$-Dy$^{3+}$, Nd$^{3+}$-Fe$^{3+}$ evidently play a major role in driving the spin-reorientation of Fe moments at low temperatures. 
\section{Discussion}
\label{discuss}
In NdFeO$_{3}$, the anisotropic nature of Nd$^{3+}$-Fe$^{3+}$ exchange interactions causes the ${\Gamma}_{4}$${\rightarrow}$${\Gamma}_{2}$ reorientation of the Fe$^{3+}$ spins in the temperature range 200-105\,K. 
In NDFO, these interactions are greatly diluted due to the Dy$^{3+}$ ions which causes the reorientation to start below 75\,K. 
In the reorientation region, effective molecular fields arise in the $a$-$c$ plane and act on the magnetic moments of Fe$^{3+}$ and the rare-earth ions. The effective field ${\bf H_{Nd-Fe}}$, arising due to the Nd$^{3+}$-Fe$^{3+}$ interactions, eventually results in ($f_{x}^R$, $c_{y}^R$) polarization of the Nd$^{3+}$ moments\cite{yamaguchi1974}. 
Due to comparatively weaker Dy$^{3+}$-Fe$^{3+}$ exchange interactions in NDFO, the extent of Dy$^{3+}$ polarization is expected to be smaller in comparison to that of the Nd$^{3+}$ moments.\par       
  Below 25\,K, the Nd$^{3+}$ moments of NdFeO$_{3}$ start to polarize in $c_{y}^R$ configuration\cite{bartolome1997single}. The similar kind of polarization also occur in NDFO. The ordering $c_{y}^R$ results in development of "$y$" component of ${\bf H_{Nd-Fe}}$, corresponding to effective field in the $b$ direction, wherein the sign is alternate on each Fe$^{3+}$ ion. 
In NDFO, the "$y$" component of ${\bf H_{Nd-Fe}}$ can induce additional reorientation of the Fe$^{3+}$ spins towards the ${\Gamma}_{4}$ phase. 
 Moreover the non-collinear calculations show that ${\Gamma}_{2}$ and ${\Gamma}_{4}$ phases are energetically very close to each other. Thus even small effective fields can result in the re-emergence of the ${\Gamma}_{4}$ magnetic structure of the Fe$^{3+}$ spins.
Thus between 20-1.5\,K, the Fe$^{3+}$ spins show an additional spin reorientation. As a result, the magnetic structure of Fe$^{3+}$ spins, in NDFO, is given by ${\Gamma}_{2+4}$(${\Gamma}_{2}$+${\Gamma}_{4}$) representation at low temperatures ($<$20\,K).\par
  In NDFO, the (100) peak associated with long range $c_{y}^R$ ordering of the rare-earth ions is observed only from 3.5\,K onwards. The same peak in NdFeO$_3$ is observed below 25\,K. The lower ordering temperature in case of NDFO might be due to lower strength of  Dy$^{3+}$-Fe$^{3+}$ interaction.  Interestingly the $R^{3+}$ ordering, in  NDFO, shows a greater magnetic moment of 1.8 ${\mu}_{B}$ indicating a partial polarization of Dy$^{3+}$ moments as well.  Enhanced magnetic moment of rare-earth ions is attributed to Nd$^{3+}$-Dy$^{3+}$ interactions.  Also, due to crystal field effects in Nd$^{3+}$ and Dy$^{3+}$  (Kramers' ions), large difference between the effective magnetic moments of both ions occurs. The Nd$^{3+}$ ion has a moment in the range $\sim$0.9-1${\mu}_{B}$, while the Dy$^{3+}$ ion has a moment of nearly 9${\mu}_{B}$\cite{bartolome1997single, Holmes1972}.
Thus the effective field can induce polarization of the large Dy$^{3+}$ moments also in the ($f_{x}^R$, $c_{y}^R$) configuration. \par
In DyFeO$_{3}$, as magnetic field along $b$ direction results in the development of ${\Gamma}_{4}$ structure, the Dy$^{3+}$ moments undergo reorientation to ($c_x^R$,$f_y^R$)\cite{Stanislavchuk2016}. However, a signature of (001) magnetic Bragg peak is not observed in NDFO, which corresponds exclusively to $c_x^R$. 
Finally, we discuss the possibility of magnetoelectric effect in NDFO. In NdFeO$_{3}$, the magnetic point group $D_{2h}$m$^{\prime}$m$^{\prime}$m$^{\prime}$ possesses inversion symmetry. As a result, its magnetoelectric tensor is zero. 
In DyFeO$_{3}$, the ${\Gamma}_{5}$($g_{x}^R$,$a_{y}^R$) magnetic structure of Dy$^{3+}$ belongs to the magnetic point group m$^{\prime}$m$^{\prime}$m$^{\prime}$ which has non-zero diagonal elements in the magnetoelectric tensor\cite{Yamaguchi1973, Zvezdin2009}. Below $T_\mathrm{N2}$, the magnetic structure of Fe$^{3+}$/Dy$^{3+}$, with ${\Gamma}_{15}$ ($A_{x}$,$G_{y}$,$C_{z}$;$g_{x}^R$,$a_{y}^R$) representation, has a polar magnetic point group ($D_{2}$)222,  which is responsible for a non-zero magnetoelectric tensor(${\alpha}_{zz}$)\cite{Yamaguchi1973}.
The magnetic point groups of NDFO, corresponding to ${\Gamma}_{2}$ and ${\Gamma}_{4}$ representation, are (C$_{2h}$)m$^\prime$m$^\prime$m and (C$_{i}$) ${\underline 2}$m$^\prime$, respectively. Both the point groups are non-polar and thus do not possess any magnetoelectric coupling terms due to inversion symmetry\cite{Yamaguchi1973}. Thus the magnetic structure of NDFO does not lead to magnetoelectric effect. From our structural studies, we do not obtain reduction in structural symmetry which can result in ferroelectric polarization. The absence of magnetoelectric effect is also confirmed from dielectric measurements in presence of magnetic field upto 5\,T (data not shown). 
%
\section{Summary and Conclusions}
\label{Summ}
To summarize, NDFO polycrystalline samples were studied in detail to understand its complex magnetic and electronic properties.
The material crystallizes in the space group $Pbnm$ with both Nd and Dy, occupying the same crystallographic position and thus being randomly distributed in the crystal. 
Though magnetic properties of NDFO are similar to that of NdFeO$_3$, there are many interesting and significant differences between the two. Below $T_\mathrm{N1}$(${\sim}$ 700 \,K), magnetic structure belongs to $\Gamma_4$($G_{x}$, $F_{z}$) representation, with the $a$ axis being the easy axis of the Fe$^{3+}$ magnetic moments. The large single anisotropy of Dy$^{3+}$ which induces an abrupt $\Gamma_4$${\rightarrow}$${\Gamma}_{1}$ spin reorientation in DyFeO$_3$ is suppressed with 50${\%}$ Nd substitution. Instead, a gradual $\Gamma_4$${\rightarrow}$${\Gamma}_{2}$ reorientation of the Fe$^{3+}$ spins is observed which begins close to 75\,K and results in magnetic structure represented by ${\Gamma}_{2}$ ($F_x$, $G_z$)  at 20 \,K with the $c$ axis as the easy axis. Interestingly, the magnetic structure given by ${\Gamma}_{4}$ re-emerges again below 20\,K.
This also coincides with the development of (100) magnetic Bragg peak which corresponds to the $c_{y}^R$-arrangement of the Nd$^{3+}$/Dy$^{3+}$ magnetic moments. This is unlike the ($G_{x}$) magnetic ordering associated usually with Dy$^{3+}$ moments. The symmetry of the rare-earth ordering also does not support any magnetoelectric coupling, which is also confirmed from field dependent dielectric measurements.
\par
At 1.5\,K, the rare-earth ordering results in magnetic moment of 1.8\,${\mu}_\mathrm{B}$ which is lower than the expected average magnetic moment of ${\sim}$5\,${\mu}_\mathrm{B}$ from Nd$^{3+}$/Dy$^{3+}$ magnetic sublattice, but the observed value (1.8\,${\mu}_\mathrm{B}$) is much higher than the experimentally observed moments from Nd-ordering in NdFeO$_{3}$. The complete absence of a ${\lambda}$-anomaly in specific heat clearly indicates that the ordering of Nd$^{3+}$/Dy$^{3+}$ moments is induced by effective molecular fields arising due to Fe$^{3+}$ spins.
The process of reorientation and rare-earth ordering is explained by our first principles density functional theory calculations, considering both collinear and non collinear spin arrangements within GGA+$U$+SO approximation. The rare-earth ordering clearly shows a preference of C-type over G-type ordering. The non-collinear spin calculations show that the Fe$^{3+}$ spins prefer to align as $G_{z}$ which is symmetry-compatible with the $c_{y}^{R}$ arrangement of the Nd/Dy moments. The various ``bare-exchange" interactions obtained for simplified unit cells, show that while the Dy$^{3+}$-Fe$^{3+}$ interaction is weakest, the Nd$^{3+}$-Fe$^{3+}$ and Nd$^{3+}$-Dy$^{3+}$ interaction compete with the Dy$^{3+}$-Dy$^{3+}$ interactions which leads to the strong polarization of the rare-earth (Nd$^{3+}$/Dy$^{3+}$) sublattice. It can be concluded that due to the highly unequal magnetic moments of the rare-earth ions, a molecular field acts on the $R$ as well as the Fe$^{3+}$ moments. The net field can lead to a successive ${\Gamma}_{2}$${\rightarrow}$${\Gamma}_{4}$ spin reorientation of the Fe$^{3+}$ spins.

\section{Acknowledgement} 
This work was supported by the UGC-DAE Consortium for Scientific Research (CSR) and Science and Engineering Research Board (SERB) through CRS-M-228, ECR/2015/000136, respectively. We acknowledge the support from IIT Roorkee through SMILE-13 grant. AS and SR acknowledge MHRD for research fellowships. MA acknowledges UGC-DAE Consortium for Scientific Research (CSR) through CRS-M-228 for research fellowships. 

\bibliography{NDFO_Neutron}
\end{document}